\documentclass[intlimits,twoside,a4paper]{article}
\usepackage{amsmath,amssymb}
\usepackage{graphicx}

\usepackage[T2A]{fontenc}
\usepackage[cp1251]{inputenc}

\usepackage[eqsecnum]{cmpj2}

\issue{2014}{17}{4}{43701}
\doinumber{10.5488/CMP.17.43701}

\title[Heterophase liquid states]%
{Heterophase liquid states: Thermodynamics, structure, dynamics%
\thanks{Dedicated to Prof. C.A. Angell on the occasion of his
$80^\textrm{th}$ birthday.}}
\author[A.S. Bakai]{A.S. Bakai}
\address{
NSC Kharkiv Institute of Physics and Technology, 61108 Kharkiv, Ukraine
}

\date{Received December 4, 2013, in final form June 17, 2014}
\authorcopyright{A.S. Bakai, 2014}

\begin{document}

\maketitle

\begin{abstract}
 An overview of theoretical results and experimental data on  the thermodynamics,
structure and dynamics of the heterophase glass-forming liquids  is presented. The
theoretical approach is based on the mesoscopic heterophase fluctuations model (HPFM)
developed within the framework of the bounded partition function approach.
The Fischer cluster phenomenon, glass transition, liquid-liquid transformations,
parametric phase diagram, cooperative dynamics and  fragility of the glass-forming
liquids is  considered.

\pacs 78.35.+c, 64.70.Pf

\end{abstract}

\section{Introduction}
\label{sec:1}
Structure of a glass-forming liquid and glass possesses a short-range and
medium-range order (SRO  and MRO) rather than a long-range order (LRO). Below the crystallization
temperature, $T_{m} $,  precautions have to be taken to avoid crystallization or
a quasi-crystalline structure formation and to prevent  the supercooled liquid state
down to the glass transition. Therefore, a liquid can be transformed into amorphous
(glassy) solid only if  cooling is fast enough to avoid crystallization. As a
result, the liquid is non-equilibrium and unstable at the glass transition.  For
this reason a description of the glass transition cannot be based on the canonic
Gibbs statistics. A palliative approach based on the bounded statistics can be formulated
as follows.

If the cooling time is much longer than the equilibration time of the liquid structure
on scale $\xi $ [let us denote this time by $\tau \left(\xi \right)$] and no significant
structural correlation occurs on scales $r>\xi $, one can consider the glass transition
as a sequence of transformations of the structure states which are equilibrated just
on the scales $r<\xi $. Statistical description of such a liquid can be developed if
we exclude from the statistics the states with the correlation scale $r>\xi $ and,
on the other hand, ensure that the observation time, $\tau _\textrm{obs} $, is much
longer than $\tau \left(\xi \right)$.  In this case, the Gibbs partition function
can be replaced by the bounded partition function which is used then to determine
the free energy of the partially equilibrated liquid.   Limitation of the phase
space due to the exclusion of the states with correlation lengths $r>\xi $ leads to an increase
of the free energy of the equilibrium state. The standard Gibbs statistics restores
with $\xi \to \infty $. The observation time limits from above the scale of the relaxation
time $\tau \left(\xi \right)$ and, consequently, the scale $\xi $, because $\tau \left(
\xi \right)$ increases $\sim\xi ^{\theta } $ (the exponent $\theta $ depends on
the features of the relaxation kinetics).

 The spatial scale of the SRO, $\xi _\textrm{SRO} $,  is minimal among the possible
correlation lengths in the liquid. Accordingly, $\tau \left(\xi _\textrm{SRO} \right)$ is
the shortest structure relaxation time because it is controlled by rearrangement
of a comparatively small number of directly interacting molecules. The formation of longer
correlations, with $\xi \gg \xi _\textrm{SRO} $, which involves a  large number
of molecules in rearrangement and is  driven by relatively weak multi-molecular forces, takes much
longer than $\tau \left(\xi _\textrm{SRO} \right)$ time. The liquid or glass, which is equilibrated
on the scale $r\sim \xi _\textrm{SRO} $ without considerable correlations on larger scales,
is the minimally ordered amorphous state which can be considered using the bounded
statistics method. For this reason, as the first step, the bounded partition
function should be considered taking into account the states with equilibrated SRO.

It is experimentally established that the glass-forming liquids are heterophase (their
structure consists of the mutually transforming fluid-like and solid-like substructures).
Observations of the heterophase structure of glass-forming liquids are numerous.
Among significant observations of the last decades we should mention the formation
of the  Fischer cluster (fractal aggregate of the solid-like HPF in glass-forming
organic liquids and polymers \cite{1,2,3,4,5,6,7,8,9,10}),  evolutive HPF  in supercooled triphenyl phosphit
observed in \cite{11}, and others\footnote{Survey article \cite{12} is devoted to the physics of heterogeneous glass-forming liquids.}.

Many types of the SRO  usually  coexist in glasses.  In  Bernal's mechanical
model of the dense random packing of hard spheres, six types of the local order (Bernal's
holes) are statistically significant and nearly one third of them are non-crystalline
\cite{13}. Similar results are obtained using computer simulations of liquids
and glasses with different interatomic potentials \cite{14,15,16,17,18}\footnote{ Just a few of
a huge number of papers devoted to this subject are cited. }.

 A wide spectrum of relaxation times in glass-forming  liquids is observed
due to the variety of SRO types (see \cite{19,19'} and references cited).

Since SRO is the molecular order formed due to microscopic forces, the correlation
length $\xi _\textrm{SRO} $ is equal to or exceeds the range of direct molecular interactions.
Therefore, to describe heterophase states, a mesoscopic theory is needed, in which
 molecular species of size $r\sim \xi _\textrm{SRO} $ specified by SRO are
``elementary'' structural elements rather than molecules.
These are not molecular potentials that determine the equilibrium states and relaxation dynamics of heterophase states but rather the parameters of heterophase fluctuation interactions connected with molecular potentials.
Evidently, the mesoscopic
Hamiltonian is more universal but less detailed than the microscopic Hamiltonian
specified by molecular potentials. Parameters of the mesoscopic Hamiltonian can
be considered as phenomenological coefficients with averaged out microscopic details
of molecular interaction.

These ideas are in the base of the heterophase fluctuation model (HPFM)  \cite{10,20,21,22,23,23',23'',24}
which is considered  in sections~\ref{sec:2}--\ref{sec:6} and in appendixes \ref{sec:A} and \ref{sec:B}. It is further  used
while considering the issues of the thermodynamics of a liquid-glass transition and
polymorphous transformations of glass-forming liquids and glasses induced by
the SRO reordering and mutual ordering of heterophase fluctuations  (section~\ref{sec:7} and appendix \ref{sec:C}).  The cooperative relaxation dynamics of a heterophase liquid is
considered within the framework of phenomenological model formulated in HPFM \cite{10,21}
(section \ref{sec:8}). Conclusive remarks are placed in section~\ref{sec:9}.

\section{Hetrophase fluctuations and the order parameter}
\label{sec:2}

 The heterophase fluctuation  is an embryo of a foreign phase in the matrix phase.
In many liquids, even in normal state (above the crystallization temperature, $T_{m}$),
solid-like species are revealed by means of difractometry. The first observation
of such heterophase fluctuations  (HPF) was made by  Stewart and Morrow  \cite{25}. They
have discovered  sybotactic groups  (transient molecular solid-like clusters possessing
specific short-range order) in simple alcohols above $T_{m}$.

The HPF are non-perturbative fluctuations in contrast to perturbative fluctuations
of physical quantities near their equilibrium values in the homophase state\footnote{Review
\cite{27} is a good introduction to the physics of HPF. The role of  non-crystalline solid
embryos in  vitrification of organic low-molecular substances (e.g., phenols) was
discussed  by Ubbelohde in \cite{28}.}.  Theory of the heterophase states originates from
Frenkel's paper \cite{26}. Frenkel has coined the term ``heterophase fluctuations'' and
explored the thermodynamics of heterophase states of fluid and gas in the vicinity
of the phase coexistence curve. Frenkel's theory is applicable to all kinds of the
coexisting phases (including the fluid and solid phases) far below the critical point.
In this case, the amount of substance belonging to HPF is small, and thus Frenkel's droplet
model, with non-interacting nuclei of a foreign phase, properly describes the heterophase
state.

Frenkel's theory fails in the case of strong HPF, when the fraction of molecules
belonging to the ``droplets'' is large (for example, when it is near or exceeds the
percolation threshold), and thus droplet-droplet interaction cannot be neglected.  Besides,
this theory was not generalized to include in its consideration the states with many
SRO-types of the nucleating ``droplets''.  Both these restrictions of the Frenkel
model are obviated in the  HPFM.

The HPFM  is based on the statistics of the transient solid-like and fluid-like mesoscopic
species (clusters) which are  called $s$-fluctuons  and $f$-fluctuons, respectively. By
definition, each fluctuon is specified by SRO. The minimal size of a fluctuon is
equal to the SRO correlation length, $\xi _\textrm{SRO} $. An arbitrary number of types of
the $s$-fluctuons, $m\geqslant 1$, can be included into consideration.

To escape needless complications, let us assume that the fluctuons are uniform-sized
with size  $r_{0} $ and with the number of molecules per fluctuon equal to $k_{0} \sim
r_{0}^{3} $. Thus, $\xi _\textrm{SRO} \simeq r_{0} $. This simplification is reasonable from
the physical point of view because in the both states SRO is formed due to the action
of the same microscopic forces, and the difference of the densities of a liquid and a solid
usually amounts to just a few percent.  The solid-like and fluid-like fractions
consist of $s$- and $f$-fluctuons, respectively.

Let us denote by $N$ the total number of molecules of liquid and by$N_{f} ,N_{1} ,\ldots,N_{m} $ the
numbers of molecules belonging to $f$- and $s$-fluctuons,
\begin{equation}
 N_{f} +N_{1} +\ldots+N_{m} =N.
 \label{GrindEQ__1_}
\end{equation}
The total number of fluctuons is $N_\textrm{fluct} =N/k_{0} $.

The $(m+1)$-component order parameter of the heterophase liquid is determined as follows:
\begin{equation}
\left\{c\right\}=(c_{f} ,c_{1} ,\ldots,c_{m} ),
\qquad
c_{i} =\frac{N_{i} }{N} \geqslant 0, \qquad i=f,1,\ldots,m \, ,
\label{GrindEQ__2_}
\end{equation}
\begin{equation}
c_{f} +c_{1} +\ldots+c_{m} \equiv c_{f} +c_{s}
=1.
\label{GrindEQ__3_}
\end{equation}
Evidently, $c_{i} $ is the probability of the molecule belonging to $i$-th type of fluctuons. $N_{s}
=N_{1} +N_{2} +\ldots+N_{m} =c_{s} N$ is the number of molecules of the solid-like fraction.
The spatial distribution of the fluctuons on scale $r\gg r_{0} $ can be described by
the order parameter fields $c_{i} (x)$ with mean values equal to $c_{i} $.

Let us regard the  $k$-th type $s$-fluctuons as statistically insignificant if  $c_{k}
\ll m^{-1} $.  The $f$-fluctuons become statistically insignificant if  $c_{f} \ll 1$.
The exclusion of the statistically insignificant components of the order parameter from
consideration allows one to simplify the equations of HPFM. The statistically insignificant
entities, when necessary, can be included into consideration as perturbations.

\section{ The quasi-equilibrium glass transition and ``ideal'' glass}
\label{sec:3}

Let us consider  more in detail the formulated in Introduction conditions under which
the glass transition with equilibrated SRO takes place:
\begin{enumerate}
\item  The liquid cooling time or the observation time, $\tau _\textrm{obs} $,  should be
less than the time of  crystallization,
\begin{equation}  \tau _\textrm{obs} \ll \tau _\textrm{LRO}\, ,
\label{GrindEQ__4_}
\end{equation}
 $\tau _\textrm{LRO} $ is the time of  long-range ordering.
\item  The observation time is much longer than the time of short-range order
equilibration,
\begin{equation}
\tau _\textrm{obs} \gg \tau _\textrm{SRO} \sim \tau _{\alpha}\,.
 \label{GrindEQ__5_}
\end{equation}
\end{enumerate}
Reordering of SRO due to localized cooperative rearrangement of the molecular structure
is an elementary $\alpha $-relaxation event. Therefore, it is put $\tau _\textrm{SRO} =
\tau _{\alpha } $ ($\tau _{\alpha } $ is the $\alpha $-relaxation time).

The condition \eqref{GrindEQ__4_} limits the value of  $\tau _\textrm{obs} $ from above.
The temperature-time-transformation diagram can be used to estimate $\tau
_\textrm{LRO} $  and to outline the area on the $(t,T)$-plane in which the condition
\eqref{GrindEQ__4_} is satisfied.

  The condition \eqref{GrindEQ__5_} restricts the value of $\tau _\textrm{obs} $ from below.
It implies that the SRO is equilibrated during the glass formation. Hence, the
order parameter \eqref{GrindEQ__3_} is a function of $P$ and $T$ and
depends on time $t$ just because $P$ and $T$ depend
on $t$. When this condition is satisfied, the glass transition can be considered
as a sequence of quasi-equilibrium transformations of the SRO.

Due to a dramatic increase of $\tau _{\alpha}$  with the temperature decrease near $T_\textrm{g} $,
the condition \eqref{GrindEQ__5_} can be satisfied just above $T_\textrm{g} $. Evidently,
the condition \eqref{GrindEQ__5_} cannot be satisfied below the temperature  $T_\textrm{F}
\left(\tau _\textrm{obs} \right)$ determined as the root of the equation
\begin{equation}
\tau _{\alpha } \left(T\right)\big|_{T_\textrm{F}
} =\tau _\textrm{obs}\,.
\label{GrindEQ__6_}
\end{equation}
 This is the temperature of kinetic glass transition because below $T_\textrm{F} \left(\tau
_\textrm{obs} \right)$ the SRO can be considered as ``frozen''.  Glass transition temperature   $T_\textrm{g} $ determined
from the viscosity measurements or by means of calorimetry or dilatometry at the
same thermal history is usually equal to $T_\textrm{F} \left(\tau _\textrm{obs} \right)$ with good
accuracy,  i.e., $T_\textrm{g} \simeq T_\textrm{F} $.

In the limiting case, with $\tau _\textrm{obs} \to \infty $ and $\tau _\textrm{obs} \ll \tau _\textrm{LRO} $,
when both conditions \eqref{GrindEQ__4_} and \eqref{GrindEQ__5_} are satisfied, the
quasi-equilibrium cooling of a liquid  leads to the formation of  hypothetical ``ideal''
glass (with equilibrated SRO and MRO but without any LRO). Hereinafter, the term ``ideal
glass'' is used in this sense.

It is worth to note that due to the condition \eqref{GrindEQ__4_}, the residual configurational
entropy of the ``ideal'' glass is not equal to zero at $T\to 0$ because any two parts
of such a glass can be considered as non-correlated and statistically independent if
the distance between them exceeds the largest correlation length which is finite
by definition.

In publications, the issues concerning the physical properties of equilibrium amorphous
states below $T_\textrm{g} $ are often debated. Between them, the hypothetical vanishing
and  non-analyticity of the configurational entropy, $S_\textrm{conf} (T)$, as a function
of temperature,  at a finite temperature $T_\textrm{K} $  (the Kauzmann paradox) \cite{29}, and
Vogel-Fulcher-Tamman  singularity of  $\tau _{\alpha } \left(T\right)$ at a temperature  $T_\textrm{VFT} $ \cite{30,31,32}
are under discussion. In the Adam-Gibbs model \cite{33}, the Kauzmann ``entropy crisis''
is included as an assumption which leads to  the VFT relaxation time singularity
at $T_\textrm{K} $. Thus, in the Adam-Gibbs model $T_\textrm{VFT} =T_\textrm{K} $. The values of $T_\textrm{K} $ and $T_\textrm{VFT} $ found
from the fittings of data on thermodynamics and dynamics of many glass-forming liquids
are close, $T_\textrm{VFT} \approx T_\textrm{K} $. Due to the above noted absence of the ``entropy
crisis'' in the ``ideal'' glass, one can conclude that  $T_\textrm{K} $ and $T_\textrm{VFT} $ should be considered as free parameters of the widely used phenomenological model \cite{33}.
The issue of proximity of   $T_\textrm{K} $ and $T_\textrm{VFT} $ is considered and confirmed within
the framework of HPFM in \cite{34}.

\section{ Mesoscopic free energy of the heterophase liquid}
\label{sec:4}

The phenomenologic free energy of the heterophase liquid in terms of the introduced
order parameter can be presented in the form of polynomial expansion in powers of  $\left
\{c_{i} \left(x\right)\right\}$,
\begin{equation}
G\left(P,T;\left\{c(x)\right\}\right)=G_\textrm{L}
\left(P,T\right)+G_\textrm{V} \left(P,T\right).
\label{GrindEQ__7_}
\end{equation}
In the summand $G_\textrm{L} (P,T)$, just local interactions of the fields $\left\{c_{i}
(x)\right\}$ are included,
\begin{equation}
G_\textrm{L} (P,T)=\int g_\textrm{L} (x,P,T)\rd^{3} x,
\label{GrindEQ__8_}
\end{equation}
\begin{eqnarray}
g_\textrm{L} (x,P,T) &=&\sum _{i}c_{i}(x)g_{i}^{0} \left(P,T\right)+
\frac{z}{2} \sum _{i,k}c_{i}(x)c_{k} (x)g_{ik}^{0} \left(P,T\right)  \nonumber \\
&&  + T\sum
_{i}c_{i}(x)\ln c_{i} (x) +g_{0}\left(P,T\right).
\label{GrindEQ__9_}
\end{eqnarray}
$g_{i}^{0}\left(P,T\right)$ is independent of the order parameter free energy of $i$-th
fluctuon; $g_{ik}^{0} \left(P,T\right)\geqslant 0$ is the fluctuonic pair interfacial free
energy; $z$ is the fluctuonic coordination number which is taken as independent of
the fluctuon type.

The summand $G_\textrm{V} (P,T)$ describes contribution of non-local (volumetric) interaction
of $s$-fluctuons, which is taken in the following form
\begin{equation}
G_\textrm{V} (P,T)=\frac{N}{k_{0} } 2\pi \sum _{i,j}
\int \Phi (r)w_{ij} (r)r^{2} \rd r, \qquad
r=\left|\vec{x}-\vec{x}'\right|,
\label{GrindEQ__10_}
\end{equation}
\begin{equation}
w_{ij} (r)=\langle c_{i} (x)c_{j} (x')\rangle =V^{-1}
\int c_{i} (x)c_{j} (x') \rd^{3} x,
\label{GrindEQ__11_}
\end{equation}
$w_{ij} (r)$ is the pair correlation function of $s$-fluctuons, $V$ is the volume, $\Phi
(r)$ is the potential of pair interaction of the s-fluctuons. This interaction, analogous
to the attraction potential of colloid particles in a solvent, plays a significant
role in states with diluted solid-like species because it provides aggregation of
the $s$-fluctuons, leading to the Fischer cluster formation. It is taken as Yukawa
potential with cutoff range $R_{0}$ which is larger than but comparable
with $r_{0} $,
\begin{equation}
\Phi (r)=-\frac{\varphi }{r} \exp (-r/R_{0}).
\label{GrindEQ__12_}
\end{equation}
Fluctuonic short-range correlation appears due to both local and volumetric interactions.
The Ornstein-Zernike (OZ) equation \cite{35} can be used to estimate the fluctuonic correlation
length, $\xi _{fl} $.  It follows from OZ equation that far from a critical point,  $\xi
_{fl} $ is  comparable with the correlation length of the direct correlation function,
which, in turn, is comparable with the range of the fluctuonic pair interaction potential.
With  $R_{0} \leqslant 2r_{0} $ we have that $\xi _{fl} \simeq 2r_{0} \simeq 2\xi _\textrm{SRO} $.
As it is seen, the ordering of  fluctuons causes extension of the molecular pair correlations
beyond $r_{0} $ and the formation of the of molecular MRO. The liquid region of size $\xi
_{fl} $ with correlated fluctuons is referred to as correlated domain.

The fact that the components of the order parameter $A_{i} \left(x\right)$ are normalized
probabilities, which cannot exceed 1, validates the presentation of  $G\left(P,T
\right)$ in the form of the  polynomial expansion in powers of  $\left\{c_{i} \left(x
\right)\right\}$.

The connection of the phenomenological free energy (\ref{GrindEQ__7_})--(\ref{GrindEQ__12_}) with the Gibbs free energy
can be found using the approach formulated in \cite{36}. It is shown \cite{36}  that the free
energy presented in terms of the order parameter plays the role of the efficient Hamiltonian
in the Gibbs statistics and determines the most probable state of the system. The interplay
between the mesoscopic free energy and the Gibbs statistics is considered in
appendix~\ref{sec:A}.

\section{The fluctuon-fluctuon  interaction and the frustration parameter}
\label{sec:5}

The physical meaning of the pair interaction coefficients of the neighboring fluid-like
and solid-like fluctuons is clear. It is the fluid-solid interfacial free energy
taking into account the geometry of the contacting fluctuons.

The solid-like fraction can be considered as a mosaic composed of $s$-fluctuons with
different SRO. The interfacial free energy of a pair of $s$-fluctuons depends on their
mutual orientations. Evidently, coherent joints of the non-crystalline $s$-fluctuons
is hampered at any orientation. The interfacial free energy increase due to the geometric
badness of the fit of contacting $s$-fluctuons is \textit{the structural frustration parameter}\footnote{ For
more information on the structural frustration see e.g., \cite{37} and references cited.
The importance of the frustration parameter at glass transition was considered and discussed
qualitatively in \cite{38,38'}.  A  specific frustration parameter avoiding the critical point
is introduced in the model of frustration-limited domains (FLD) \cite{39,40}.}.    Because
of its importance, let us consider the fluctuonic frustration parameter  more in detail.

A non-crystalline solid-like cluster grows due to the attachment of new molecules. Hence,
the former surface molecules become the inner ones and the non-crystalline
cluster structure becomes frustrated because not all newly formed coordination polyhedra
are exactly similar to the initial polyhedron. A part of them can have the geometry
similar to that of the initial coordination polyhedron but slightly deformed. The occurrence
of the coordination polyhedra of completely different geometry is also possible.
Thus, if the initial coordination polyhedron has some symmetry, the newly formed
coordination polyhedra have a violated or completely changed symmetry. Consequently,
the binding energies of the attached molecules appear smaller than that of the inner
molecule.

A decrease of the binding energy per molecule is accompanied by an increase of the
configurational entropy due to ambiguities of the geometrical changes of the new
coordination polyhedra.

As an example, let us consider the growth of a $z$-vertex coordination polyhedron in
the case when the addition of a new coordination shell leads to the formation of $z-1$
new coordination polyhedra with similar but deformed initial coordination polyhedron
while one of them has a different geometry. In this case, the energy of the inner
$z+1$ molecules is
\begin{equation}
E_{z+1} =\varepsilon _{0} \left(z+1\right)+\bar{\varepsilon }_\textrm{def}
(z-1)+\varepsilon _{1} =\varepsilon _{0} \left(z+1\right)+\varepsilon _\textrm{frust}\,,
\label{GrindEQ__13_}
\end{equation}
 $\varepsilon _{0} $ is the mean energy of the initial cluster, $\bar{\varepsilon
}_\textrm{def} $ is the mean energy of deformation and $\varepsilon _{1} $ is the energy
of a molecule with the coordination polyhedron of different geometry. The last two
terms in r.h.s. of (\ref{GrindEQ__13_}) determine the frustration energy, $\varepsilon
_\textrm{frust} $. Because of uncertainty of the last molecule position, the frustration
configurational entropy due to this uncertainty is as follows:
\begin{equation}
s_\textrm{frust} =s_{z+1} =k_\textrm{B}^{} \ln z.
\label{GrindEQ__14_}
\end{equation}
The frustration free energy is as follows:
\begin{equation}
g_\textrm{frust} =\varepsilon _\textrm{frust} -Ts_\textrm{frust}\,.
\label{GrindEQ__15_}
\end{equation}
As it is seen, $\varepsilon _\textrm{frust} $ is $\sim z$  while $s_\textrm{frust} \sim
\ln z$. Therefore, $g_\textrm{frust} >0$ with  $z\gg1$.

One can conclude that generally the structure of interfacial layer of contacting
fluctuons is frustrated and that $g_\textrm{frust} >0$.

\section{ Equations of the liquid state equilibrium }
\label{sec:6}

Variation of the free energy functional (\ref{GrindEQ__4_}) at condition (\ref{GrindEQ__3_})
yields the equations of equilibrium state,
\begin{equation}
\frac{\delta }{\delta c_{i} (x)} G(P,T)+\lambda
\frac{\partial }{\partial c_{i} (x)} \sum _{k}c_{k} (x) =0,
\label{GrindEQ__16_}
\end{equation}
$\lambda $ is the Lagrange multiplier.

Let us denote by $\mu _{i} (P,T)$ the derivative
\begin{eqnarray}
\mu _{i} (P,T)&=&\frac{\partial \left(g_{l} +g_{v} \right)}{\partial c_{i} }=g_{i}^{0}
+\sum _{k}c_{k} (x) g_{ik} +T\ln c_{i}(x) \nonumber \\
&&
+\sum _{j}\int \Phi (x,x')c_{j} (x')\rd^{3}
x'.
\label{GrindEQ__17_}
\end{eqnarray}
Here, $g_{ik} =zg_{ik}^{0} $. Variables $\left(P,T\right)$ are not shown.

As a result, it follows from (\ref{GrindEQ__16_}) that
\begin{equation}
\mu _{f} (P,T)=\mu _{1} (P,T)=\ldots=\mu _{m}
(P,T)=-\lambda.
\label{GrindEQ__18_}
\end{equation}
These equations are analogous to the Gibbs equations of the equilibrium of phases.

Equilibrium
state is stable if the quadratic form $\left\| \frac{\delta ^{2} G}{\delta c_{i}
\delta c_{k} } \right\| $ is positively definite.

\section{Solutions of the equations of state}
\label{sec:7}

\subsection{Two-state approximation}
\label{sec:7.1}

In the physics of glass-forming liquids, different two-state models are in use for
a long time \cite{39, 41,42,43,44,44',45,46,47,47',48,48',48'',49}. HPFM in the two-state approximation provides abbreviated
entry of the glass transition.

In fact, in the two-state approximation of the HPFM, the mesoscopic substructure
of the solid-like fraction is neglected and the order parameter in the two-state approximation
has just two components,  $c_{s} $ and $c_{f} $,
\begin{equation}
c_{s} +c_{f} =1.
\label{GrindEQ__19_}
\end{equation}
Applying the spatial averaging, we obtain from (\ref{GrindEQ__17_})--(\ref{GrindEQ__18_})
\begin{equation}
 \left(1-2c_{s} \right)\tilde{g}_{sf} +T\ln \frac{c_{s} }{1-c_{s}
} =h_{sf}\,.
\label{GrindEQ__20_}
\end{equation}
Here,
\begin{equation}
\tilde{g}_{sf} =g_{sf} -g_{ss} /2; h_{sf} =g_{f}^{0}
-g_{s}^{0} -g_{ss} /2,
 \label{GrindEQ__21_}
\end{equation}
\begin{equation}
g_{s}^{0} =\sum _{k}c_{k}^{*}  g_{k}^{0} +T
\sum _{k}c_{k}^{*}  \ln c_{k}^{*}, \qquad     g_{ss} =\sum _{}g_{ik}  c_{i}^{*} c_{k}^{*}\,,
\qquad
c_{i}^{*} =c_{i} /c_{s}\,,
 \label{GrindEQ__22_}
\end{equation}
$g_{ss}$ is the frustration parameter. It depends on the interaction coefficients
of the $s$-fluctuons and probabilities $\left\{c_{i}^{*} \right\}$. For a while, the
volumetric interactions (\ref{GrindEQ__12_}) are not accounted for.

In the two-state approximation, the coefficient $g_{sf} $ and the frustration parameter $g_{ss} $ are
taken as constants. Some remarks concerning the accuracy of two-state approximation
of HPFM  appear in section~\ref{sec:9}.

Equation (\ref{GrindEQ__20_}) is isomorphic to the equation of state of the Ising
model with an external field $h_{sf} $.  The solution of equation (\ref{GrindEQ__20_}) at $c_{s}\ll 1$ is as follows:
\begin{equation}
c_{s} (T)=\exp \left\{ \left[\Delta s_{fs} \left(T_{e}^{0}
\right)\left(T_{e}^{0} -T\right)-g_{sf} \right]/T_{e}^{0} \right\}.
\label{GrindEQ__23_}
\end{equation}
Here,
\begin{equation}
\Delta s_{f,s} \left(T\right)=-\frac{\partial \left(g_{s}^{0} -g_{f}^{0}
\right)}{\partial T} =s_{f} \left(T\right)-s_{s} \left(T\right)
\label{GrindEQ__24_}
\end{equation}
is the difference of entropies of the $f$- and $s$-fluctuon. $T_{e}^{0}$ is the solution of the equation
\begin{equation}
g_{f}^{0} \left(P,T_{e}^{0} \right)=g_{s}^{0}
\left(P,T_{e}^{0} \right).
 \label{GrindEQ__25_}
\end{equation}
At $c_{f} =1-c_{s} \ll 1$
\begin{equation}
c_{f} (T)=\exp \left\{ \left[\Delta s_{f,s} \left(T_{e}^{1}
\right)\left(T_{e}^{1} -T\right)-g_{sf} \right]/T_{e}^{1} \right\},
\label{GrindEQ__26_}
\end{equation}
where $T_{e}^{1} $ is the solution of the equation
\begin{equation}
g_{f}^{0}\left(P,T_{e}^{1} \right)=g_{s}^{0} \left(P,T_{e}^{1} \right)+g_{ss}\,.
\label{GrindEQ__27_}
\end{equation}
The physical meaning of the characteristic temperatures $T_{e}^{0}$, $T_{e}^{1}$ is explained
below.

\begin{figure}[!b]
\centerline{
\includegraphics[width=.42\textwidth]{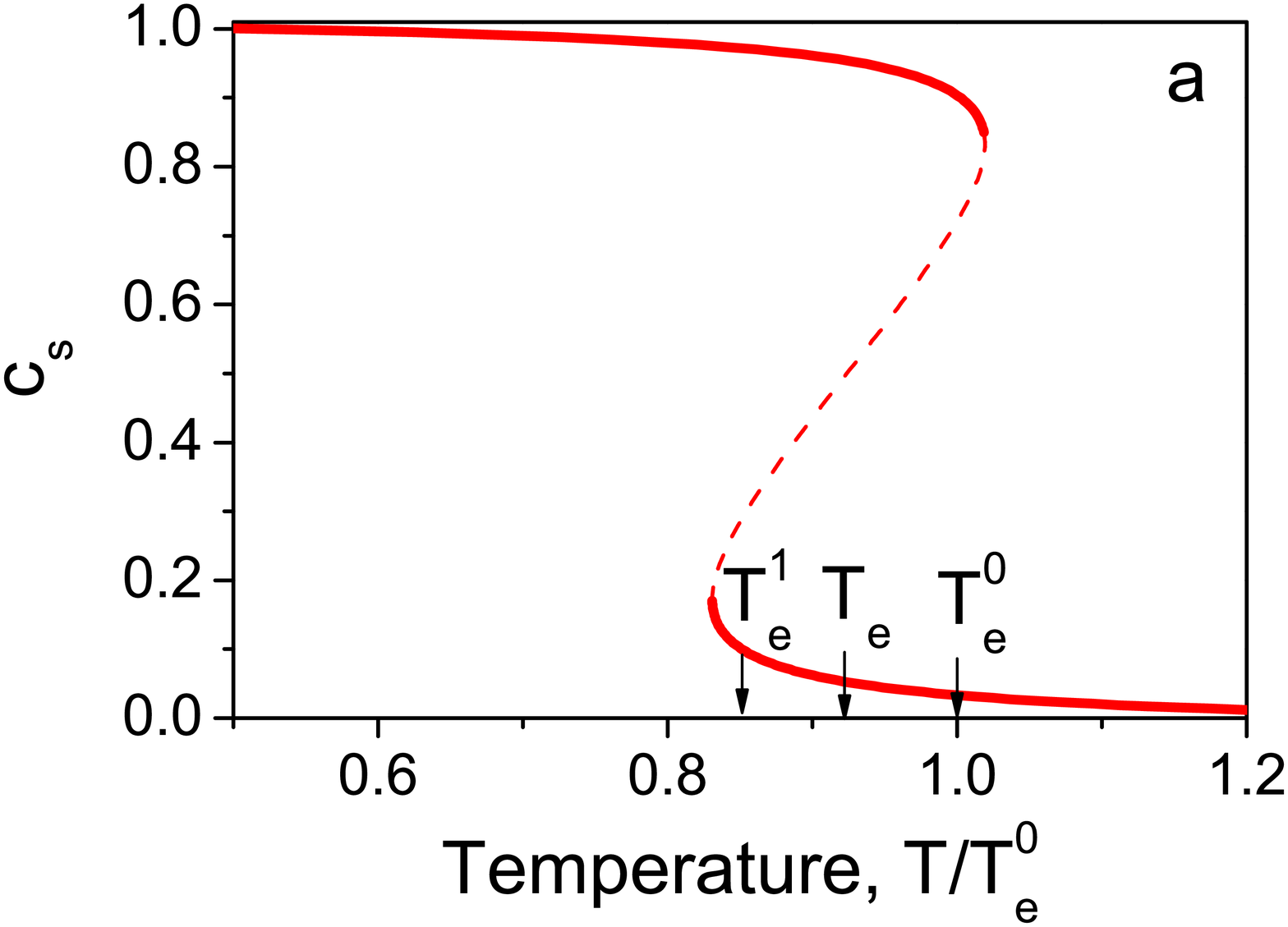}
\hspace{5mm}
\includegraphics[width=.4\textwidth]{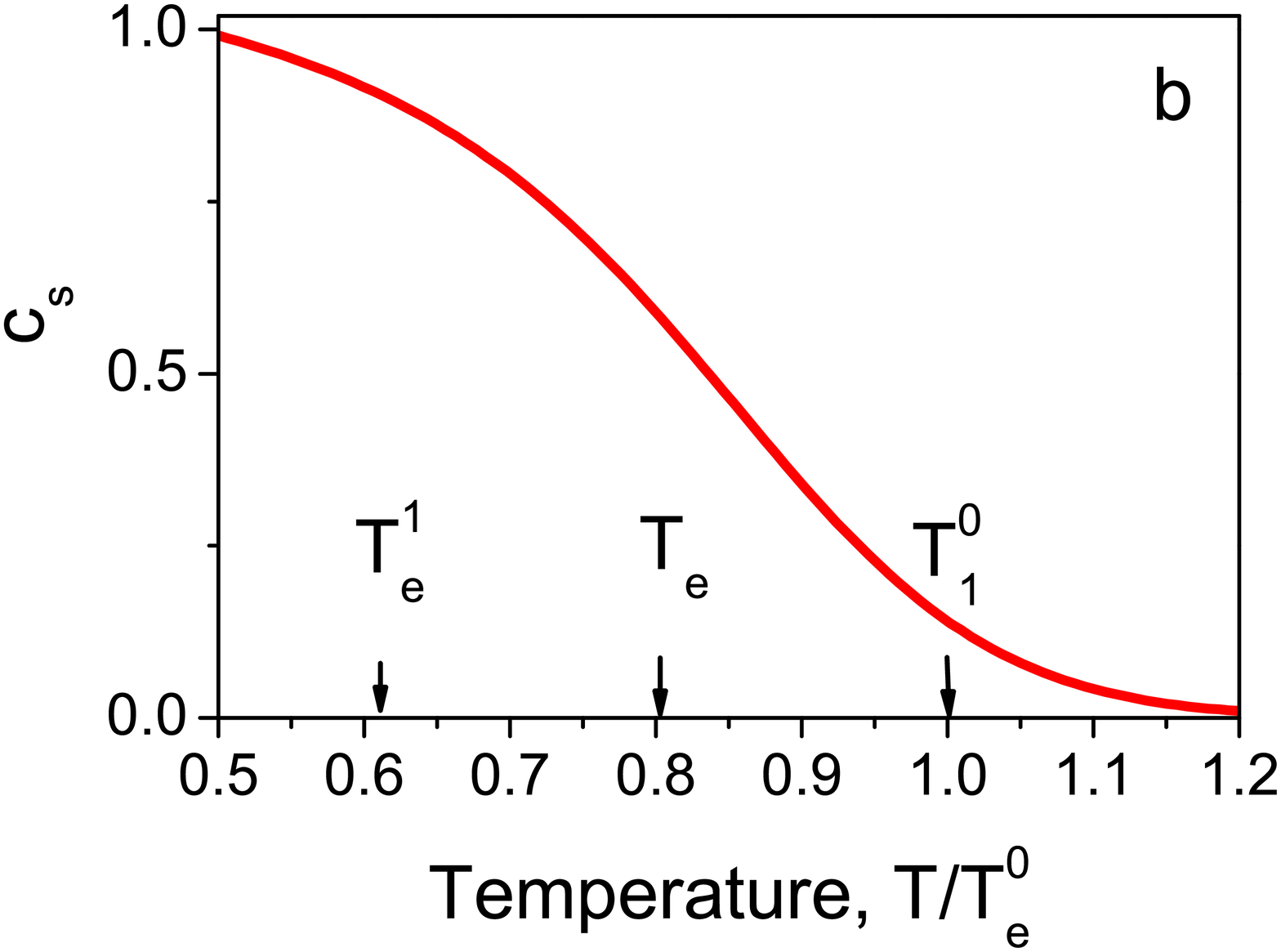}
}
\caption{\label{fig1} (Color online)
The solid-like fraction of of  liquid, $c_{s} $, vs $T/T_e^0 $ at (a) $\tilde
{g}_{sf} \left( {P,T_e } \right)>2T_e $ and (b) $\tilde {g}_{sf} \left(
{P,T_e } \right)<2T_e $.
}
\end{figure}

 The temperature $T_{e} $, at which the ``external field'' $h_{sf} $ is equal to
zero,  is the coexistence temperature of  two heterophase liquid states determined
by equation
\begin{equation}
g_{f}^{0} \left(P,T_{e} \right)=g_{s}^{0} \left(P,T_{e}
\right)+g_{ss} /2.
\label{GrindEQ__28_}
\end{equation}
At $T=T_{e} $, we have  $c_{s} (T_{e} )=c_{f} (T_{e} )=1/2$. In the vicinity of  $T_{e} $,
\begin{eqnarray}
c_{s} \approx \frac{1}{2} +\frac{h_{sf} (T)}{2(2T_{e} -\tilde{g}_{sf}
)} \left[1-\frac{2Th_{sf}^{2} (T)}{3(2T_{e} -\tilde{g}_{sf} )^{3} } \right]
=\frac{1}{2}
+\frac{\Delta s_{s,f} (T_{e} )\left(T-T_{e} \right)}{2(2T_{e} -\tilde{g}_{sf} )} +O
\left((T-T_{e} )^{3} \right).
\label{GrindEQ__29_}
\end{eqnarray}
As it follows from (\ref{GrindEQ__25_}), (\ref{GrindEQ__27_}) and (\ref{GrindEQ__28_}),
\begin{equation}
T_{e}^{0} \approx T_{e} +g_{ss} /2\Delta s_{f,s}\,,
\qquad
T_{e}^{1} \approx T_{e} -g_{ss} /2\Delta s_{f,s}\,.
\label{GrindEQ__30_}
\end{equation}
The solution (\ref{GrindEQ__29_}) is stable at $\tilde{g}_{sf} \left(P,T_{e} \right)<2T_{e} $.
If  $\tilde{g}_{sf} \left(P,T_{e} \right)>2T_{e} $, it is unstable and at $T=T_{e},
(P)$ the first order phase transition takes place.

Graphic representation of  solutions of equation (\ref{GrindEQ__20_}) is shown in figure~\ref{fig1}. The stable and unstable solutions are depicted by solid and dashed lines, respectively.

If $\tilde{g}_{sf}=2T_{e} $, i.e.
\begin{equation}
g_{ss} =2g_{sf} -4T_{e}\,,
\label{GrindEQ__31_}
\end{equation}
then, the 2nd order phase transition takes place at $T=T_{e} (P)$.
In accordance with (\ref{GrindEQ__23_}) and (\ref{GrindEQ__26_}), above $T_{e}^{0} $ and
below $T_{e}^{1} $,  the HPF are weak but  within the temperature range $\left[T_{e}^{1}
,T_{e}^{0} \right]$, where in compliance with (\ref{GrindEQ__29_}) $c_{s} $ and $c_{f} $ are
comparable quantities, they are strong.

\begin{figure}[!t]
\centerline{ \includegraphics[width=.3\textwidth]{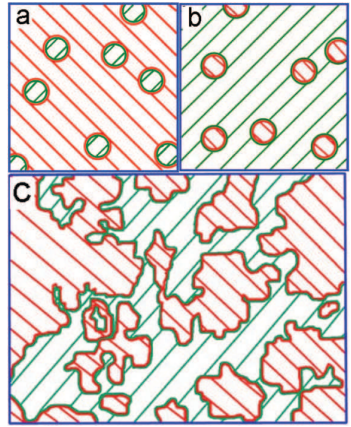} }
\caption{\label{fig2}
(Color online) Schematic representation of the hetrophase liquid states: (a) rare $s$-fluctuons
in fluid  described by equation (\ref{GrindEQ__23_}); (b) $f$-fluctuons in glass [equation (\ref{GrindEQ__26_})];
(c) heterophase state with comparable fractions of the $s$-  and $f$-fluctuons [equation (\ref{GrindEQ__29_})].
}
\end{figure}
\begin{figure}[!b]
\centerline{ \includegraphics[width=.35\textwidth]{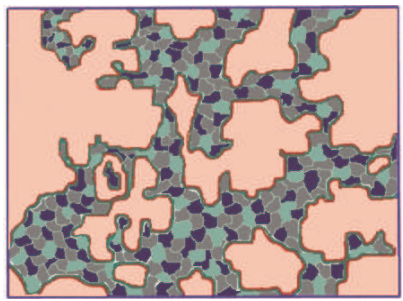} }
\caption{\label{fig3} (Color online)
The same as in figure~2~(c) but the mesoscopic structure of the solid-like fraction
containing several types of the $s$-fluctuons is shown.
}
\end{figure}

It worth to note that the solutions (\ref{GrindEQ__23_}) and (\ref{GrindEQ__26_})
reproduce the results of  Frenkel's model in the vicinity of the phase coexistence temperatures
(here, $T_{e}^{0} $ and  $T_{e}^{1} $, respectively). Therefore, $T_{e}^{0} $ can be
considered as the coexistence temperature of the fluid and heterophase liquid phases
while  $T_{e}^{1} $ is the phase coexistence temperature of the ``ideal'' glass (as
it is determined above) and the heterophase liquid. Thus, $T_{e}^{1} $ is the ideal
glass transition temperature.  The real glass transition  temperature, $T_\textrm{g} $,
which depends on $\tau _\textrm{obs} $ (see section~\ref{sec:3}), is above  $T_{e}^{1} $ due to
dramatic retarding of the structure relaxation with temperature decrease. For this
reason, the real glass transition temperature range,  $\left[T_\textrm{g} ,T_{e}^{0} \right]$, is
narrower than $\left[T_{e}^{1} , T_{e}^{0} \right]$.
The structure
of the heterophase states in the vicinity of the characteristic temperatures $T_{e}^{1} $, $T_{e} $ and $T_{e}^{0} $ is
schematically presented in figure~\ref{fig2}.
In figure~\ref{fig3}, the mesoscopic structure of the solid-like fraction with several types
of $s$-fluctuons is shown schematically. Let us remind that the solutions of equations (\ref{GrindEQ__23_}),
(\ref{GrindEQ__26_}), (\ref{GrindEQ__29_}) are obtained under the assumption that the
fractions of $s$-fluctuons $\left\{c_{i} \right\}, i=1,\ldots,m$ are nearly
constant or they are changing continuously and smoothly. This assumption fails if
a phase transformation with stepwise changes of the fractions $\left\{c_{i} \right
\}$ within the solid-like fraction takes place.  In the next section, the impact of such
a phase transformation within the solid-like fraction on the features of the fluid-solid
phase transformation is considered.

\subsection{ Phase transition in the solid-like fraction}
\label{sec:7.2}

Evidently, a phase transition in the solid-like fraction causes a non-analytic behaviour
of the solutions of equation \eqref{GrindEQ__20_}. This type of the liquid-liquid transition
appears due to multiplicity and interaction of the $s$-fluctuons which leads to the
mutual ordering and phase separations within the solid-like fraction.

As a minimal model, let us consider the heterophase liquid with two types of $s$-fluctuons.
Hence,  $m=2$. Thus, in \eqref{GrindEQ__18_} $i,j=1,2$.  The equation of state \eqref{GrindEQ__18_} for the solid-like
fraction is as follows:
\[
\left(1-2c_{1}^{*} \right)c_{s} \tilde{g}_{12} +T\ln \frac{c_{1}^{*} }{1-c_{1}^{*}
} =h_{12}\,,
\qquad
c_{i}^{*} =c_{i} /c_{s}\,,
\qquad
c_{1}^{*} +c_{2}^{*} =1\,,
\]
\begin{equation}
\tilde{g}_{12} =g_{12} -\left(g_{12} +g_{22}
\right)/2,
\qquad
h_{12} =g_{2}^{0} -g_{1}^{0} +c_{s} \left(g_{22} -g_{11} \right)/2.
\label{GrindEQ__32_}
\end{equation}

It
is seen that this equation is isomorphic to equation \eqref{GrindEQ__20_} but the  ``external
field''  $h_{12} $ and the pair interaction coefficient $c_{s} \tilde{g}_{12} $
depend on $c_{s} $. Therefore, associated  solutions of equations \eqref{GrindEQ__20_}
and \eqref{GrindEQ__32_} should be considered together. The search for a general solution
of these nonlinear equations at an arbitrary set of coefficients is a cumbersome and
hardly attractive task because the values of the coefficients for substances are
initially unknown. Nevertheless, we can look for some ``typical'' solutions at a reasonable
specification of the coefficients.

As a useful example, let us consider solutions of equation \eqref{GrindEQ__32_} in the
vicinity of the coexistence curve, $h_{12} \left(P,T\right)=0$, assuming that $c_{s}
\left(g_{22} -g_{11} \right)$ is a negligible quantity. In this case, the coexistence
temperature, $T_{12} $, is determined by equation
\begin{equation}
g_{1}^{0} \left(P,T_{12} \right)=g_{2}^{0}
\left(P,T_{12} \right).
\label{GrindEQ__33_}
\end{equation}

\begin{figure}[!b]
\centerline{
\includegraphics[width=.38\textwidth]{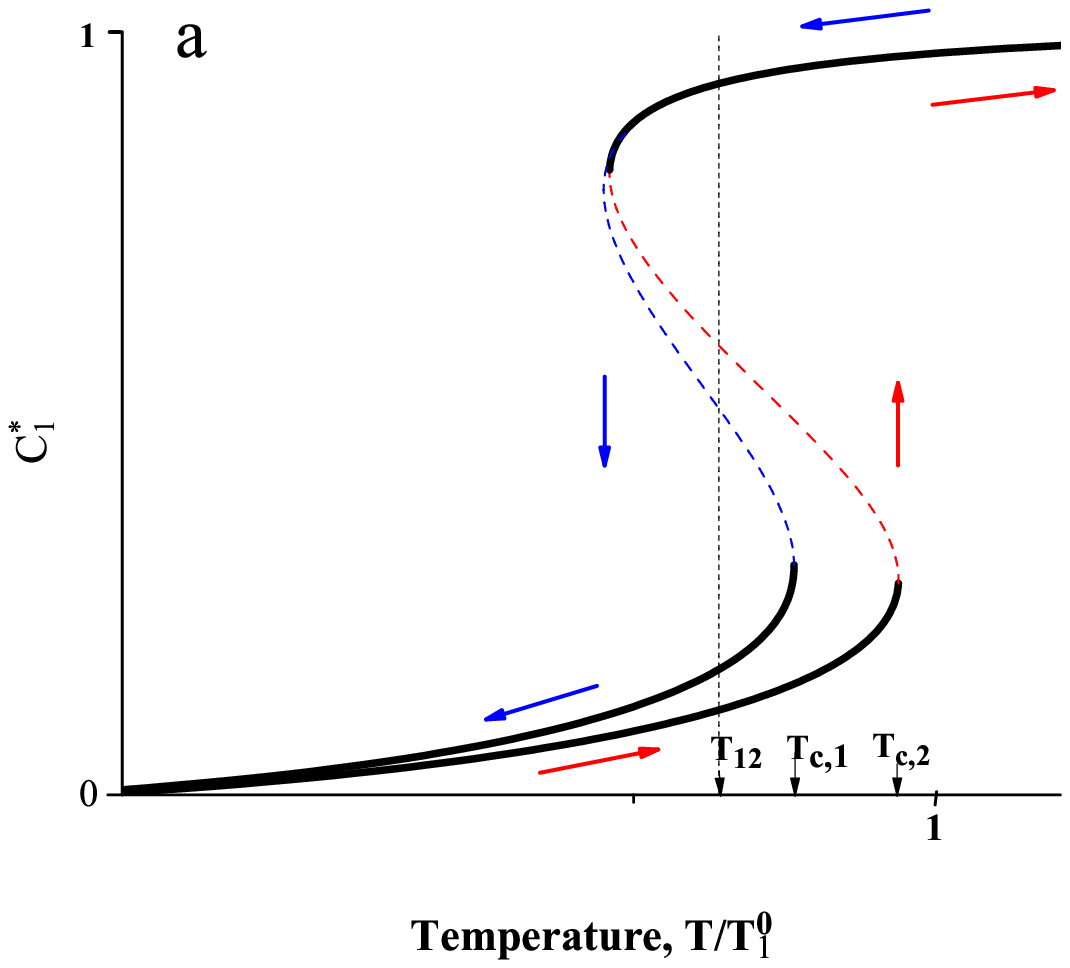}
\hspace{5mm}
\includegraphics[width=.41\textwidth]{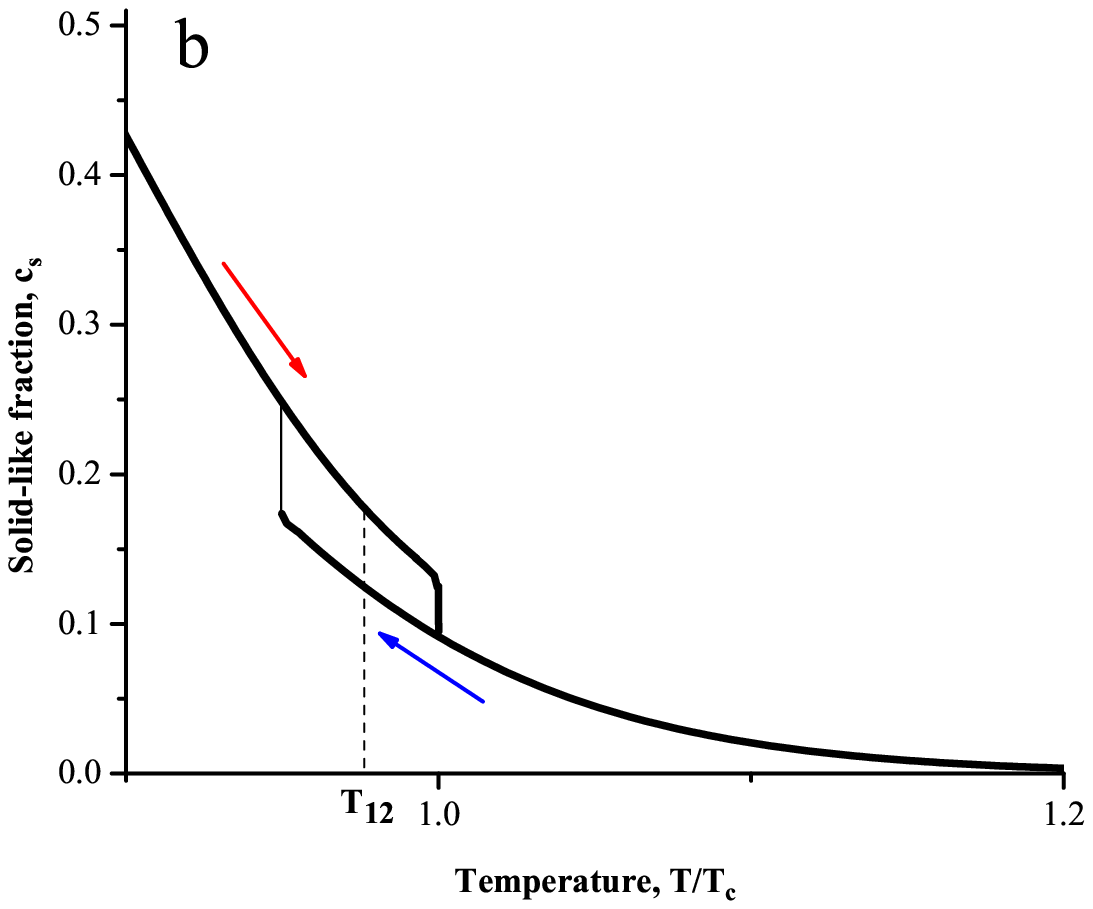}
}
\caption{\label{fig4}
(Color online) First order phase transition within the solid-like fraction (a) induces the
liquid-liquid first order phase transition (b).
}
\end{figure}

It is assumed that $T_{12} $ is above the coexistence temperature $T_{e} $.  A phase
transformation of the solid-like fraction and induced liquid-liquid phase transition
at $T_{12} <T_{e} $ is considered in \cite{24}.

In the vicinity of $T_{12} $
\begin{equation}  h_{12} =g_{1}^{0} \left(P,T\right)-g_{2}^{0}
\left(P,T\right)\approx \left(s_{1} -s_{2} \right)\left(T-T_{12} \right)\equiv \Delta
s_{12} {\kern 1pt} \left(T-T_{12} \right),
\label{GrindEQ__34_}
\end{equation}
$s_{1} $ and $s_{2} $ is the entropy of  $s$-fluctuon of type 1 and 2, respectively.

If $c_{s}\left(T_{12} \right)\tilde{g}_{12} >2T_{12} $, then there are two stable solutions
of equation \eqref{GrindEQ__32_} and, as a consequence, two stable solutions of equations  \eqref{GrindEQ__20_},
\eqref{GrindEQ__21_}. Consideration of these solutions is contained in appendix~\ref{sec:B}.
Their graphic representation is shown in figure~\ref{fig4}.  The jump of the  parameter $c_{s} $ at $T=T_{12} $ is
[see (\ref{B.7}) in appendix~\ref{sec:B}] as follows:
\begin{equation}
\Delta c_{s} \left(T_{12} \right)\approx \left(s_{1}
-s_{2} \right)\exp \left(-g_{sf} \beta \right).
 \label{GrindEQ__35_}
\end{equation}

\begin{figure}[!t]
\centerline{
\includegraphics[width=.39\textwidth]{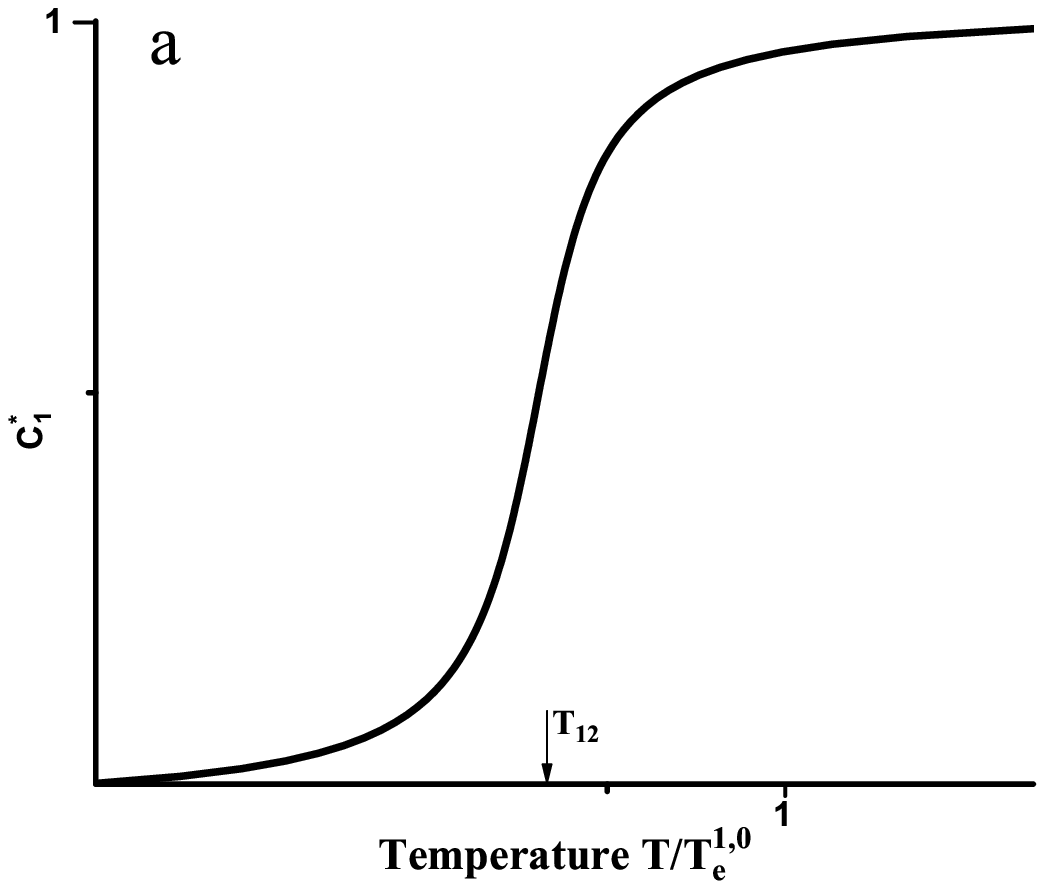}
\hspace{5mm}
\includegraphics[width=.41\textwidth]{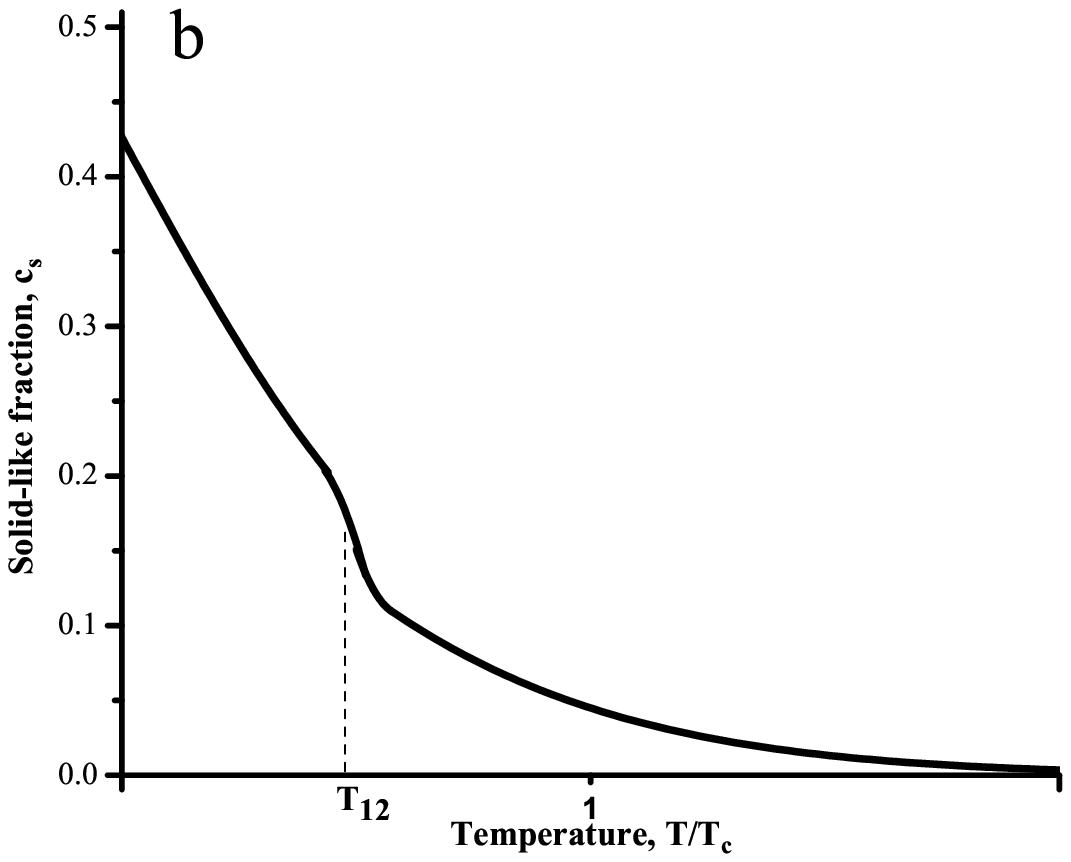}
}
\caption{\label{fig5}
Continuous but stepwise evolution of  $c_{1}^{*} \left(T\right)$shown in  a)
induces the appearance of the inflection point of  $c_{s} \left(T\right)$ at  $T=T_{12} $,
b).
}
\end{figure}
If $c_{s} \left(T_{12} \right)\tilde{g}_{12} <2T_{12} $, the fractions $c_{1}^{*}
,c_{2}^{*} $  and parameter $c_{s} $ change continuously at $T=T_{12} $ but
the  inflection points of  the functions $c_{1}^{*} \left(T\right)$  and  $c_{s}
\left(T\right)$ appear  at $T=T_{12} $ (figure~\ref{fig5}). Hence, the expression \eqref{GrindEQ__35_}
estimates the bench height of the parameter $c_{s} $.

The heat of the phase transitions is determined by equation (\ref{B.8}) in appendix~\ref{sec:B},
\begin{equation}
\Delta H_{12} \simeq c_{s} \left(T_{12} \right)(s_{1}-s_{2})k_{0}^{-1}\left[H_{fs}+T_{12}\right],
\label{GrindEQ__36_}
\end{equation}
$H_{fs} $ is the heat of the fluid-solid phase transition. It is taken into account here that the heat of  $1\leftrightarrow 2$
solid-solid phase transition
is equal to $(s_{1}-s_{2})k_{0}^{-1}T_{12}$.

It is worth to note that at a fixed value of  $c_{s} $, equation \eqref{GrindEQ__32_} is isomorphic
to Ising model with non-zero external field and with exchange integral $c_{s} \tilde{g}_{12} $ which
can be positive or negative. The external field controls the ratio $c{}_{1}^{*} /c{}_{2}^{*} $ while
the sign of the exchange integral determines the type of mutual ordering of the $s$-fluctuons.
With $c_{s} \tilde{g}_{12} >0$ (ferromagnetic interaction), $s$-fluctuons of different
types tend to separate. At $c_{s} \tilde{g}_{12} <0$, the ``antiferromagnetic''
order with alternating $s$-fluctuons of different types is preferable. In both  cases,
the fluctuonic SRO  generates molecular medium-range order with the correlation length $\xi
_{fl} \sim 2r_{0} $ [in compliance with the general conclusion made in section~\ref{sec:4} after
equation (\ref{GrindEQ__12_})].

\subsection{The Fischer cluster }
\label{sec:7.3}

\begin{figure}[!t]
\centerline{
\includegraphics[width=.4\textwidth]{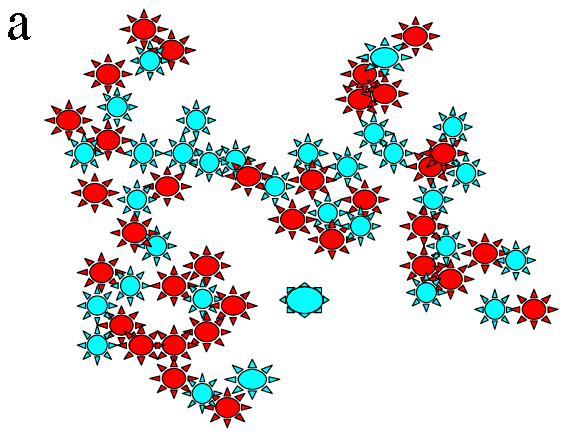}
\hspace{5mm}
\includegraphics[width=.4\textwidth]{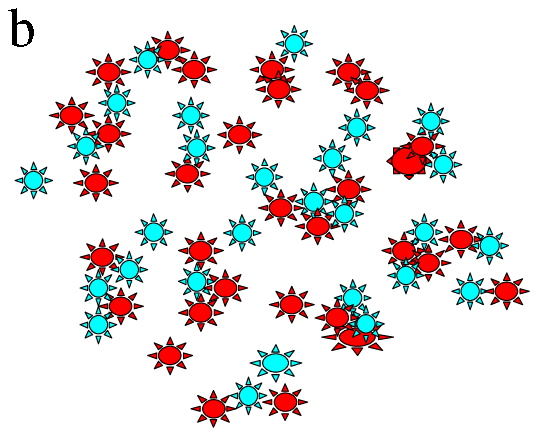}
}
\caption{\label{fig6}
(Color online) Schematic fragments of the heterophase liquid with (a) and without (b)  Fischer's
cluster. Just two types of statistically significant $s$-fluctuons are shown.  The
circle shows the size of the correlated domain.
}
\end{figure}

Along with the above  considered types of MRO appearing due to local fluctuonic interaction,
there is a different type of the fluctuonic order with comparatively large  (as it
was observed, up to $\sim 102$~nm) correlation length, $\xi _\textrm{FC} \gg\xi _{fl} $. It
appears due to the aggregation of the $s$-fluctuons under the effect of the volumetric gravitation
potential \eqref{GrindEQ__12_}. The equilibrated aggregation of $s$-fluctuons possesses
the fractal structure with fractal dimension, correlation length, equilibration time
and relaxation dynamics depending on the liquid features and temperature. This remarkable
phenomenon, which was discovered and investigated in detail by Fischer et al. \cite{1,2,3,4,5,6,7,8,9,10,22,23,23',23'',24},
is known as the Fischer cluster.  The Fischer cluster was visualized by observing
a speckle pattern in ortho-terphenil  \cite{6}. The speckle pattern fluctuates and rearranges
very slowly, with characteristic time $\sim 1$~min at $T=293$~K, while  the $\alpha $-relaxation
time at this temperature is  $\tau _{\alpha } =40$~ns. With temperature increase, the
speckle size and contrast decreases and at $T>340$~K  no speckle is seen. Schematically,
the heterophase liquid structure with and without the Fischer cluster is shown in
figure~\ref{fig6}. It is worth to note that the Fischer cluster formation in heterophase liquid
is not an exclusion but the rule if the Fischer cluster equilibration time, $\tau _\textrm{FC} $,
is shorter than the observation time, i.e., if
$\tau _{\alpha } \ll \tau _\textrm{FC} \ll \tau _\textrm{obs} \ll \tau _\textrm{LRO} $.  The heterogeneous structure
and slow structural relaxation are observed  not only in many Van der Waals molecular
liquids but also in some  metallic melts above  $T_{m} $  \cite{50,51,52}.

The Fischer cluster was originally identified using the results of the small-angle
X-ray scattering on the density fluctuations. The conventional large-scale density
fluctuations in a homophase liquid are proportional to the isothermal compressibility $\kappa
_{T} $ and are independent of  the wave vector $q$,
\begin{equation}
\left|\rho (q)\right|^{2} \left. \right|_{q
\to 0} \sim \kappa _{T} T.
 \label{GrindEQ__37_}
\end{equation}
Here, $q$ is the wave vector, $\rho (q)$ is the amplitude of density fluctuations.
The intensity of X-ray scattering on the density fluctuations, $I\left(q\right)$, is
proportional to $\left|\rho (q)\right|^{2} $.

It appears that a $q$-dependent excess scattering intensity, $I_\textrm{exc} \left(q
\right)\sim q^{-D} $ ($D$ is the fractal dimension) occurs
at $T<T_{A} \approx T_{e}^{0} $. The $I_\textrm{exc} \left(q\right)$ is much larger
than the scattering intensity on the thermal fluctuations \eqref{GrindEQ__37_}.  The results
of the wide-angle X-ray scattering show that SRO of the liquid contains both the
fluid-like and solid-like components at $T<T_{A} $ \cite{9,53}. It turns out that the thermodynamics
and $\alpha $-relaxation dynamics are quite the same in the liquid states with and
without the Fischer cluster. It means that the changes of the thermodynamic properties
due to the Fischer cluster formation are too small to be reliably detected and that
the fluctuonic SRO does not undergo noticeable changes. Therefore, the Fischer cluster
formation can be considered as the process of self-organization of the correlated domains
(CDs), i.e., entities possessing the fluctuonic SRO with the correlation length $\xi _{fl}
\approx 2r_{0} $, (section~\ref{sec:4}).

Theory of the Fischer cluster is developed in \cite{10,22,23,23',23''}.  The Fischer cluster is
considered as a fractal aggregation of $s$-fluctuons with the fractal dimension \textit{$D_{f} $} and
correlation length $\xi _\textrm{FC} $. Minimization of the free energy as function of $\bar{c}_{s} $, $D_{f} $ and $\xi
_\textrm{FC} $ allows one to determine the equilibrium values of $D_{f} $ and $\xi _\textrm{FC} $.
It is found (see appendix~\ref{sec:C}) that
\begin{equation}
\xi _\textrm{FC} \left(\overline{c}_{s} ,D_{f} \right)
\approx \left(\overline{c}_{s} \right)^{\frac{1}{D_{f} -3} } r_{0}\,,      \qquad       D_{f}
=3-\left(\ln \frac{r_{0} }{\xi _{fl} } \right)^{-1} \ln c_{s,\textrm{CD}}\,,
\label{GrindEQ__38_}
\end{equation}
$c_{s,\textrm{CD}} $ is
the concentration of  $s$-fluctuons within CD (\cite{23,23',23''}, appendix~\ref{sec:C})
\begin{equation}
c_{s,\textrm{CD}} =\left[1+\frac{I(D_{f} ,\xi _\textrm{FC}
)}{\partial ^{2} g^{0} \left(\overline{c}_{s} \right)/\partial \overline{c}_{s}^{2}
-I(D_{f} ,\xi _\textrm{FC} )} \right]\overline{c}_{s} \geqslant \overline{c}_{s}\,.
\label{GrindEQ__39_}
\end{equation}
The
quantity $I(D_{f} ,\xi )$, (\ref{C.21}),  is proportional to $\phi _{0} $. Equation \eqref{GrindEQ__39_}
shows that within the CD, concentration of $s$-fluctuons is larger than its mean value $c_{s} $: $c_{s,\textrm{CD}}
\simeq c_{s} +\textrm{const}\,\phi _{0} $.

The liquid state with the Fischer cluster is stable (while the state without the Fischer
cluster is metastable or unstable) at
\begin{equation}
c_{s} >c_{s,0} <\left(\frac{r_{0} }{\xi _{fl}
} \right)^{2} \approx 0.16, \qquad      1<D_{f} \leqslant 3.
\label{GrindEQ__40_}
\end{equation}

Transformation of the state without Fischer's cluster into the state with Fischer's cluster
is a weak first order phase transition with the transformation heat $\propto \varphi
_{0} $.

The upper bound of the $c_{s} $-range in which the Fischer cluster exists,  $c_{s,1}
=(\xi _{fl} /r_{0} )^{2} c_{s,0} $ [see (\ref{C.26})], decreases, $\sim \phi _{0} $, with
an increase of the strength of the  $s$-fluctuons gravitation potential $\phi _{0} $.
When $c_{s} $ approaches $c_{s,1} <1$ from below, the fractal dimension $D_{f}
{\kern 1pt} {\kern 1pt} $ approaches 3 and $\xi _\textrm{FC}\to \infty $. It means that at $c_{s} \geqslant c_{s,1} $ the solid-like fraction consists
of the connected 3-dimensional solid-like clusters of size $r>\xi _{fl} $.  Thus, at $c_{s}
\geqslant c_{s,1} $, the topology of the heterophase
liquid equilibrated on scale $r\geqslant \xi _{fl} $ changes.

\subsection{Parametric phase diagram}
\label{sec:7.4}

In the HPFM, the structure and phase states of the heterophase liquid are described
in terms of $T$ and coefficients $g_{sf}$, $g_{ss}$, $h_{sf} $. It is useful to
construct a phase diagram (\textit{the parametric phase diagram}) of the glass-forming
liquid in terms of these parameters\footnote{ Tentative phase diagrams of  glass-forming
liquid in terms of the  model coefficients are introduced in \cite{38} and then in \cite{54}.}.  The parametric phase diagram of the two-state approximation is determined by equations~\eqref{GrindEQ__25_}, \eqref{GrindEQ__27_}, \eqref{GrindEQ__28_} and the equation~\eqref{GrindEQ__40_}
in combination with \eqref{GrindEQ__23_}. Namely, they determine the coexistence
temperatures of different states in terms of the coefficients $g_{sf} ,g_{ss} ,h_{sf} $.
The quantities  $T_{e}^{0} ,T_{e}^{1} {\kern 1pt} ,T_{e} {\kern 1pt} {\kern 1pt}
{\kern 1pt} {\kern 1pt} {\kern 1pt} $and $T_\textrm{FC} $ are coexistence temperatures
of
\begin{enumerate}
\item  the fluid and heterophase liquid ($T_{e}^{0} $);

\item  the heterophase liquid and ``ideal'' glass($T_{e}^{1}$);

\item  the fluid-like and solid-like states ($T_{e} $);

\item  the heterophase liquid with and without the Fischer cluster ($T_\textrm{FC} $).
\end{enumerate}
Introducing the scaled temperature, $T^{*} =T/T_{e}^{0}  $,
and the frustration parameter $g_{ss}^{*} =g_{ss} /\Delta s_{f,s} T_{e}^{0} $,
we can present the relations \eqref{GrindEQ__26_} in a dimensionless form,
\begin{equation}
T_{e}^{0*} =1,     \qquad
T_{e}^{*} =1-g_{ss}^{*} /2,  \qquad T_{e}^{1*} \approx 1-g_{ss}^{*}\,.
\label{GrindEQ__41_}
\end{equation}
The
end critical point location on the fluid-solid phase coexistence curve, $T^{*} =T_{e}^{*}
\left(P\right)$ is located at
\begin{equation}
g_{ss,\textrm{c}}^{*} ={2g_{sf} \mathord{\left/ {\vphantom
{2g_{sf}  \left(\Delta s_{s,f} T_{e}^{0} \right)}} \right. \kern-\nulldelimiterspace}
\left(\Delta s_{s,f} T_{e}^{0} \right)} -4T_{e}^{*} \left(P\right)>0.
\label{GrindEQ__42_}
\end{equation}
The
first order fluid-solid phase transition on the phase coexistence curve takes place
at $g_{ss}^{*} <g_{ss,\textrm{c}}^{*} $.

\begin{figure}[!t]
\centerline{ \includegraphics[width=.45\textwidth]{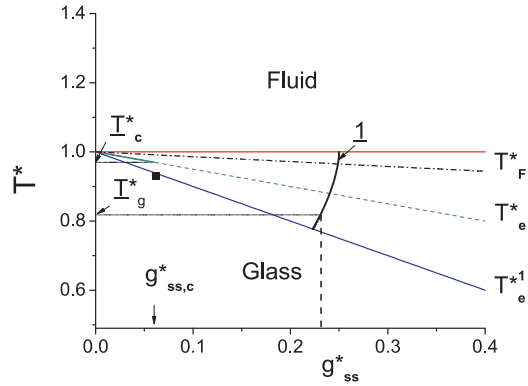} }
\caption{\label{fig7}
(Color online) Parametric phase diagram on the plane $\left(T^*,g^*_{ss} \right)$, $T^*=T/T_{e}^{0} $,  $g^*_{ss}
=g_{ss} /\Delta s_{s,f} T_{e}^{0} $. The phase coexistence lines $T=T_{e}^{0}$; $T=T_{e}$; $T=T_{e}^{1}$ and
the threshold of the Fischer cluster formation
temperature, $T_\textrm{FC}^{*} $, approximately determined using equations \eqref{GrindEQ__23_},
\eqref{GrindEQ__40_} are shown. The critical end point at $g_{ss,\textrm{c}}^{*}$ exists on
the line $T=T_{e} $. Bold line 1 schematically presents an evolution phase curve
of the equilibrium system.
}
\end{figure}

The parametric phase diagram depicted on the plane $\left(T^{*} ,g_{ss}^{*} \right)$  using
relations \eqref{GrindEQ__41_}, \eqref{GrindEQ__42_} and \eqref{GrindEQ__40_}, \eqref{GrindEQ__23_}
is shown in figure~\ref{fig7}. As an example, here is also shown one phase trajectory which
becomes non-physical below the glass transition temperature $T_\textrm{g}^{*} $. Within
the range $0<g_{ss}^{*} <g_{ss,\textrm{c}}^{*} $ the first order phase transition takes place
on the phase coexistence line $T^{*} =T_{e}^{*} $. A weak first order phase transition
takes place on the line  $T^{*} =T_\textrm{FC}^{*} $.

\subsection{Static structure factor and the order parameter restoration }
\label{sec:7.5}

\begin{figure}[!b]
\centerline{ \includegraphics[width=.45\textwidth]{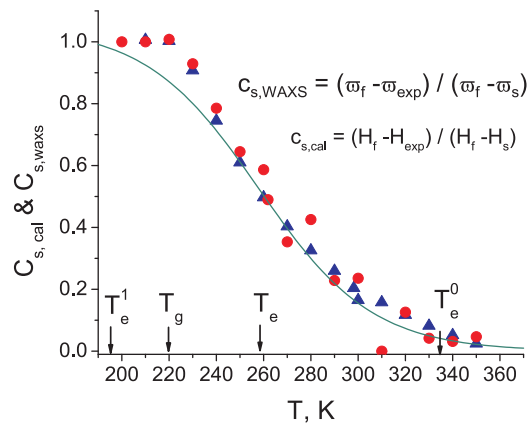} }
\caption{\label{fig8}
(Colo online) The solid-like fraction of salol vs $T$ as it is found from the analysis
of the calorimetric data (triangles), and  from the temperature dependence of the
structure factor (circles) \cite{21,53}. Arrows indicate the temperatures $T_{e}^{0}$, $T_{e}$, $T_\textrm{g}$, $T_{e}^{1}$. Line presents an analytic solution of the equation of state
in the two-state approximation with fitting parameter $\tilde{g}_{sf} $.
}
\end{figure}

Pair correlation function of the density fluctuations , $\stackrel{\frown}{\varpi
}(q,T)$,  of the heterophase liquid with the Fischer cluster is $\sim q^{-D} $ at   $r_{0}
\ll q^{-1} \ll \xi ^{-1} $.  At  $qr_{0} \sim 1$, it is a superposition of the pair correlation
functions of fluctuons,
\begin{eqnarray}
\stackrel{\frown}{\varpi }(q,T)=c_{f} (T)\varpi _{f} (q)+
\sum _{i=1}^{m}c_{i} (T) \varpi _{i} (q)
=\left[1-c_{s} (T)\right]\varpi _{f} (q)+c_{s}
(T)\varpi _{s} (q),
\label{GrindEQ__43_}
\end{eqnarray}
\begin{equation}
\varpi _{s} (q)=\frac{1}{c_{s} } \sum _{i=1}^{m}c_{i}
\varpi _{i} (q),
 \label{GrindEQ__44_}
\end{equation}
$\varpi _{f} (q)$, $\varpi _{i} (q)$ are Fourier transforms of the pair correlation
functions of the $f$- and $s$-fluctuons, respectively. The cross-correlation terms  $\sim c_{f}
c_{s} $ are omitted in \eqref{GrindEQ__43_}. The quantities $\varpi _{f} (q)$, $\varpi
_{i} (q)$ weakly depend on the temperature. For this reason, the equation \eqref{GrindEQ__43_}
can be applied to restore the order parameter $c_{s} (T)$ using the structure   factors $\varpi
_{f} (q)$, $\varpi _{s} (q)$ measured in the liquid, fluid and glassy states \cite{53}.
On the other hand, as it is shown in \cite{21}, $c_{s} (T)$ can be restored from calorimetric
data using the relation
\begin{equation}
c_{s} \simeq \left[H_{f} (T)-H_{\exp } (T)
\right]/\left[H_{f} (T)-H_{s} (T)\right]
\label{GrindEQ__45_}
\end{equation}
which follows from equation \eqref{GrindEQ__9_}. Here, $H_{f} (T),H_{s} (T)$ are enthalpies
of the fluid and glass extrapolated in the temperature range $\left[T_{e}^{1} ,T_{e}^{0}
\right]$, and $H_{\exp } (T)$is the experimentally measured enthalpy of the glass-forming
liquid. Comparison of the results of the order parameter restoration from the structural
data, using equation \eqref{GrindEQ__43_}, and from the calorimetric data, using relation
\eqref{GrindEQ__45_}, gives a good chance to check the reliability of the HPFM. This procedure
was performed using structural and calorimetric data of salol \cite{9,53}. Results are
presented in figure~\ref{fig8} by scattered symbols. Solution of the equation of state in the
two-state approximation (subsection~\ref{sec:7.1}), in which the experimentally measured thermodynamic
parameters and free parameter $\tilde{g}_{sf} $ are used, is presented there by a
solid line.

Let us remind that the analytic solution describes the order parameter $c_{s} (T)$ of
the equilibrated system. Therefore, it noticeably deviates from the experimentally
determined values $c_{s} (T)$ near the glass transition temperature, where the liquid
becomes non-equilibrium.   Relations \eqref{GrindEQ__43_} and \eqref{GrindEQ__45_},
obtained without the assumption that the system is equilibrated, allow us to recover the
thermal history of ``true'' (in the phenomenological sense) value of $c_{s} $.

\section{Dynamics}
\label{sec:8}

\subsection{$\alpha$-relaxation}
\label{sec:8.1}

Thermally activated cooperative structural rearrangements which can involve
up to $\sim 10^{2}$ molecules \cite{19,19',55,56,57,58,59}  are called  $\alpha $-relaxation. A large amount
of the molecules are involved in the rearrangement due to correlations. Structural
rearrangement of a fluctuon also involves  rearrangements of  the neighboring fluctuons
within CD  of size $\xi _{fl} $. Therefore, the size of cooperatively rearranging
domain is nearly equal to $\xi _{fl} $.

The activation energy of $\alpha $-relaxation,
\begin{equation}
E_\textrm{ac} =\frac{d\ln \tau _{\alpha } }{d\beta
}\,, \qquad \beta =\frac{1}{T}\,,
\label{GrindEQ__46_}
\end{equation}
depends on the order parameter.  It can be presented as an expansion in powers of
the order parameter \cite{10,21}\footnote{ There is no reason to believe that  $E_\textrm{ac}
\left(c_{s} \right)$  is a singular function at  $c_{s} \in \left[0,1\right]$.},
\begin{equation}
 E_\textrm{ac} =E_\textrm{ac}^{0} +E_\textrm{ac}^{1} c_{s} +E_\textrm{ac}^{2} c_{s}^{2} +\ldots \; .
 \label{GrindEQ__47_}
 \end{equation}
Above $T_{A} $, the
activation energy  is nearly equal to $E_\textrm{ac}^{0} $. Cooperativity of the liquid
dynamics is induced by the $s$-fluctuons interaction which becomes considerable below $T_{A} $.

Fischer and Bakai \cite{21} have suggested that CD can be rearranged when all the molecules therein
are in fluid-like state with correlations destroyed on the scale $\xi _{fl} $. This
assumption leads to the following expression \cite{21}
\begin{equation}
E_\textrm{ac} =\frac{A}{\left(1-T_\textrm{K} /T\right)^{2}
} +z_\textrm{CD} c_{s} \left(H_{f} -H_{s} \right)+O\left(c_{s}^{2} \right).
 \label{GrindEQ__48_}
\end{equation}
$z_\textrm{CD}
\sim \left(\xi _{fl} /a\right)^{3} $ is the cooperativity parameter, which is the mean
n umber of molecules within the CD; $H_{f}$, $H_{s} $ is the enthalpy of liquid-like
and solid-like fraction  per molecule. The first term is taken in the form proposed
for random packings of spheres in \cite{60,61}. Its denominator takes into account the
decrease of the free volume of the fluid and the numerator is equal to the activation
energy above $T_{A} $. The Kauzmann temperature, $T_\textrm{K} $, is a fitting parameter
(see comments concerning $T_\textrm{K} $ in section~\ref{sec:3})

\begin{figure}[!h]
\centerline{ \includegraphics[width=.35\textwidth,height=.38\textwidth]{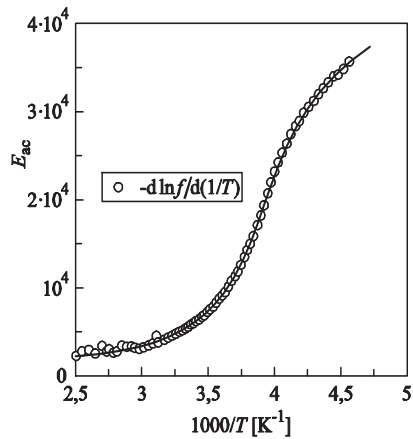} }
\caption{\label{fig9}
The activation energy of salol vs the reciprocal temperature \cite{2}.
}
\end{figure}

Enthalpies  $H_{f} \left(T\right)$  and  $H_{s} \left(T\right)$ within the temperature
range $\left[T_\textrm{g} ,T_{A} \right]$  are understood as extrapolations of these functions
measured at $T>T_{A} $  and $T<T_\textrm{g}$, respectively.

As
an example of using the equation \eqref{GrindEQ__48_} \cite{21}, the activation energy
of salol was analyzed  in \cite{21}. The activation energy of salol vs the reciprocal
temperature is shown in figure~\ref{fig9}. The experimental data are shown by circles. The curve
is a result of fitting the formula \eqref{GrindEQ__48_} using parameters $A = 967$~K; $T_\textrm{K} =153$~K, $T_{e} =257$~K; $T_{A}
\approx 325$~K; $z_\textrm{CD}= 32$; $k_{0}= 7$. The input of the first term of r.h.s of \eqref{GrindEQ__48_}
in the activation energy is nearly 10\% at $T=T_\textrm{g} $.

It is noteworthy that since the main input in the activation energy makes the term
proportional to $c_{s} $, $E_\textrm{ac} \left(T\right)$ has the inflection point at $T
\approx T_{e} $. Stickel has proposed an efficient method of analysing the $\alpha $-dynamics
to check the applicability of the Vogel-Fulcher-Tamman formula and other phenomenologic
and empiric expressions proposed for $\tau _{\alpha } (T)$ \cite{62,63,64}. He has analyzed
many molecular liquids and found the flex points of $E_\textrm{ac} \left(T\right)$ identified
as $T_{e} $. Coincidence of the values of $T_{A} $ and $T_{e} $ extracted from the
dynamic, calorimetric and structural data (figures~\ref{fig8},~\ref{fig9}) support the adequacy of  HPFM.

\subsection{Ultra-slow modes and the Fischer cluster equilibration time}
\label{sec:8.2}

Two relaxational modes are connected with the Fischer cluster. The phase transformation
of a liquid without the Fischer cluster into the state with the cluster is controlled
by nucleation and growth of a new phase. Since the phase transformation heat is small,
the phase equilibration time is rather large compared with the time of elementary
cooperative rearrangement $\tau _{\alpha } $. The Fischer cluster equilibration time
is determined in \cite{10},
\begin{equation}
\tau _\textrm{FC} \sim \frac{T}{\varphi } \left(\frac{
\xi _\textrm{FC} }{\xi _{fl} } \right)^{D_{f} +2} \frac{\tau _{\alpha } }{D_{f} -1}
\label{GrindEQ__49_}\,.
\end{equation}
 Rearrangements
of the equilibrated Fischer cluster on the scales $\xi _{fl}^{-1} \gg q>\xi _\textrm{FC}^{-1} $  are
registered as the ultra-slow modes \cite{2,3,4,5,6,7,8,9}. The relaxational rate of the ultra-slow
mode is $\sim q^{2} $. It is found within the framework of HPFM \cite{10} that
\begin{equation}
\Gamma _\textrm{usv} \left(q\right)\sim \left(q\xi
_\textrm{FC} \right)^{2} \left(\frac{\pi \xi _{fl} }{\xi _\textrm{FC} } \right)^{2} \tau _{\alpha
}^{-1} +O\left(q^{4} \right).
\label{GrindEQ__50_}
\end{equation}
The characteristic times  $\Gamma _\textrm{usv}^{-1} $ and $\tau _\textrm{FC} $ are proportional
to $\tau _{\alpha } $ and the proportionality coefficients are rather large at $\xi
_\textrm{FC} \gg \xi _{fl} $.  As it is seen,  $\tau _\textrm{FC} \gg \Gamma _\textrm{usv}^{-1} \gg \tau
_{\alpha } $.  Relations \eqref{GrindEQ__49_},\eqref{GrindEQ__50_} are in harmony
with the experimental data.

\subsection{Fragility}
\label{sec:8.3}

The fragility parameter, $\stackrel{\frown}{m}$, introduced by Angell in \cite{65,66},
is an important characteristic   of the glass-forming liquid dynamics near $T_\textrm{g} $.
It is taken as the measure of deviation of the temperature dependence of $\tau _{
\alpha } $ from the Arrhenius law. There exist strong liquids, with small fragility
parameter, $\stackrel{\frown}{m} \; \sim 10$, the most fragile liquids, with $\stackrel{
\frown}{m}\; \approx 10^{2} $, and liquids with moderate fragility in between. The
fragility parameter is tightly connected with the structural properties and thermodynamics
of a liquid, and for this reason, it is widely used at analysing the glass transition
and classification of liquids. Angell's definition of this parameter is as follows:
\begin{equation}
 \stackrel{\frown}{m}=T_\textrm{g}^{-1} \left[\frac{\rd\left(\log \tau
_{\alpha } \right)}{\rd\beta } \right]=\left. \frac{E_\textrm{ac} }{T_\textrm{g} \ln 10} \right|_{T_\textrm{g}
}  .
\label{GrindEQ__51_}
\end{equation}
In HPFM, the quantity $E_\textrm{ac} $ is determined by equation~\eqref{GrindEQ__48_}.  As it
follows from \eqref{GrindEQ__48_} and \eqref{GrindEQ__51_},
\begin{equation}
\stackrel{\frown}{m}=\frac{1}{T_\textrm{g} } \left[
\frac{A}{\left(1-T_\textrm{K} /T_\textrm{g} \right)^{2} } +z_\textrm{CD} \left[H_{f} (T_\textrm{g} )-H_{s} (T_\textrm{g}
)\right]k_{0}^{-1} c_{s} \left(T_\textrm{g} \right)\right],
\label{GrindEQ__52_}
\end{equation}
\[
H_{f} \left(T_\textrm{g} \right)-H_{s} \left(T_\textrm{g} \right)\sim \left(s_{f} -s_{s} \right)
\left(T_\textrm{g} -T_{e} \right).
\]
The main contribution in $\stackrel{\frown}{m}$ gives the second term within the
brackets of \eqref{GrindEQ__52_}. It is proportional to the number of molecules involved
in the cooperative rearrangement, $z_\textrm{CD} \sim \xi _{fl}^{3} $, as well as to the difference
of the configurational entropies of the solid-like and fluid-like species. The difference
of their vibrational entropies is comparatively small. Since these quantities can
be measured regardless of $\tau _{\alpha } $, equation~\eqref{GrindEQ__52_} permits to
check the relevance of the HPFM predictions. For example, it was found that for salol,
HPFM gives $\stackrel{\frown}{m}\;\approx 67$ \cite{10}. The fragility parameter of salol
estimated in \cite{67} is equal to 63.

As it was noted above, (subsection~\ref{sec:8.3}), the  $H_{f} \left(T\right)$  and  $H_{s} \left(T
\right)$ within the temperature range $\left[T_\textrm{g} ,T_{A} \right]$  are understood
as extrapolations of these functions measured at $T>T_{A} $  and $T<T_\textrm{g}$, respectively. Naturally, the linear or
quadratic extrapolation provides an acceptable result if the function is smooth and the higher
derivatives are small. Phase transformations in the solid-like fraction lead
to stepwise changes of $H_{s} \left(T\right)$and, consequently, to the stepwise behavior
of $\tau _{\alpha } \left(T\right)$.  In this case, extrapolations of  $\tau _{\alpha
} \left(T\right)$, determined by equation~\eqref{GrindEQ__46_}, from high and low temperatures
into the range $\left[T_\textrm{g} ,T_{A} \right]$ cannot be properly fitted. Equations~\eqref{GrindEQ__35_},
\eqref{GrindEQ__36_}, (\ref{GrindEQ__46_})--(\ref{GrindEQ__48_}) determine the temperature dependence of $\tau _{\alpha
} \left(T\right)$ in this case.

 In a series of experiments with some metallic glasses \cite{68,69,70,71,72}, the fragility parameter
value determined using the data on $\tau _{\alpha } \left(T\right)$  in the vicinity
of  $T_\textrm{g}$ and  its value recovered
from the extrapolated curve $\tau _{\alpha } \left(T
\right)$ measured at high temperatures  are completely different. As it is revealed \cite{68}, such a behavior of $\tau
_{\alpha } \left(T\right)$ of Zr-based alloy Vitreloy 4  is connected with the liquid-liquid
first order phase transition. In others melts, a  transition of this type is assumed.

Equations
\eqref{GrindEQ__35_}, \eqref{GrindEQ__36_}, (\ref{GrindEQ__46_})--(\ref{GrindEQ__48_}) provide theoretical description
of this phenomenon known as the fragile-to-strong liquid transformation. More in
detail it is  considered in \cite{73}.

\section{Concluding remarks }
\label{sec:9}

The statistical basics of HPFM include substantiation of the mesoscopic efficient
Hamiltonian and  the application of the bounded statistics method while describing
supercooled liquid states. Solutions of the equations of state of HPFM ascertain
interplay of the thermodynamic, structural and dynamic properties of the glass-forming
liquids. Thus, juxtaposing the theoretical predictions with experimental
data (as an example, see  figures~\ref{fig8},~\ref{fig9}) permits to cross-verify the adequacy of  HPFM.

The conditions of the liquid quasi-equilibrium evolution \eqref{GrindEQ__4_}, \eqref{GrindEQ__5_}
determine the applicability range of the bounded statistics in which the states with
the crystalline order  on scale $\xi \gg \xi _{fl} $ are excluded. On the other hand,
the amorphous states with the fluctuonic order having large correlation length are
included in the statistics and the Fischer cluster is described within the framework
of HPFM. Compatibility of the conditions \eqref{GrindEQ__4_}, \eqref{GrindEQ__5_}
with the fluctuonic order equilibration on large scales should be specified.

The hierarchy of  characteristic scales of spatial correlations  (starting
from the local order and molecular size \textit{a}), $a< \xi _\textrm{SRO} <
\xi _{fl} <\xi _\textrm{FC},\ldots $  is  connected
with the hierarchy of time scales $\tau \left(a\right)< \tau _{\alpha } \left(\xi _{fl} \right)<\Gamma
_\textrm{usv}^{-1} < \tau \left(\xi _\textrm{FC}
\right)$ [$\tau \left(a\right)$ is the molecule oscillation time within the cage].
Elementary step of the nucleation and growth of the crystalline embryo is the cooperative
rearrangement on the spatial and time scales $\xi _{fl} $ and $\tau _{\alpha } $, respectively.
Therefore, the formation of the crystalline embryos with the size larger than $\xi _{fl} $
takes much longer time than the SRO equilibration time. As a result, the condition
\eqref{GrindEQ__5_} can be regarded as satisfied when the condition \eqref{GrindEQ__4_}
is fulfilled.  Hence, the crystalline species of size $\sim\xi _{fl} $  coexist
with non-crystalline species within the solid-like fraction of liquid and in glass.
As a confirmation, direct observations of the structure of metallic glasses by means
of  a high resolution field ion microscopy and transmission electron microscopy
(see figure~6.5 in \cite{20}, \cite{74,75} and references cited) reveal the coexistence of crystalline
and non-crystalline structural species with sizes of up to a few nanometers.

The Fischer cluster equilibration (along with the ultra-slow modes) is observable
only if the crystallization time is much longer than $\tau \left(\xi _\textrm{FC} \right)$.
The crystallization heat (which is the thermodynamic driving force of the crystallization)
is much larger than the heat of the Fischer cluster formation. Therefore, the Fischer
cluster can be observed just in normal liquids and in the supercooled liquids with
strongly hindered crystallization.

The amount of the solid-like fraction, $c_{s} $, determines the measure of the fluid-to
solid transformation. Due to the definitive role of $c_{s} \left(T\right)$, its description
is an important issue of the theory.  The two-state approximation is a minimal model
permitting to solve this problem considering the fluid-solid  HPF states without details
of the solid-like subsystem. Evidently, this model is satisfactory  if just one type
of the $s$-fluctuons is statistically significant or when variations of the probabilities $\left
\{c_{i}^{*} \right\}$ within the glass-transition temperature range are insignificant,
i.e., if the mesoscopic structure of the solid-like fraction does not vary considerably.
At the same time, estimation of the two-state approximation accuracy shows that it
can yield acceptable results in more general cases.

The accuracy of the two-state approximation can be estimated considering the states with
transforming $s$-fluctuons. The assumption on a smooth evolution of the coefficients of equation
\eqref{GrindEQ__20_} fails   if the phase transformations, similar to those considered
in subsection~\ref{sec:7.2}, take place within the solid-like fraction. Nevertheless, even in this
case, a stepwise jump of  $c_{s} \left(T\right)$ is comparatively small because the
entropy jump and transformation heat at the solid-solid polymorphic transformation,
as a rule, is small, $\Delta s_{12} \sim 10^{-1} $ while  $\Delta s_{fs} \sim 1$. Therefore,
one can expect that the two-state approximation is acceptable with accuracy to terms $O
\left(10^{-1} c_{s} \right)$ or even better.

 Figuratively speaking, in a general case, the glass and solid-like fraction of a liquid
is a mosaic composed by the mesoscopic species of size $\sim r_{0} $ (figure~\ref{fig3}). However, unless
until the mutual ordering of $s$-fluctuons and the impact of the mosaic details on $c_{s} $ is
beyond the scope of interests, the two-state approximation can be used.

The question whether glass is a non-equilibrated highly viscous liquid or it is a non-equilibrated
solid has rather got a conceptual sense. The thermodynamic continuity of the glass transition
permits to believe that glass is a liquid with very high viscosity and long equilibration
time. But, as a matter of fact, the glass near and  below $T_\textrm{g} $, with $c_{s} \to
1$, is solid  with statictically insignificant amount of  the fluid-like species.
Nevertheless, it flows, like a polycrystal does, due to the diffusional-viscous flow
\cite{76}.   Field-emission microscopy of metallic glasses visualizes their grainy (polycluster)
structure with sizes of grains  $\sim 102$~nm. The Coble mechanism of the plastic deformation
\cite{77,78} prevails near  $T_\textrm{g} $ in such a glass \cite{76}. The grainy structure of glass
is the result of the existence of many centers of solidification within the liquid. Therefore,
the polycluster mosaic structure of glass forms in liquids with different features
of molecular forces.  Slow structural relaxation hinders the ``reclusterization''
processes and the formation of  ``ideal'' glass.

The Fischer cluster topology changes with an increase of the solid-like fraction.
Its fractal dimension $D_{f} $ is less than 3 at $c_{s} <c_{s,1} $ and it is equal
to 3 at $c_{s} \geqslant c_{s,1} $ (see subsection~\ref{sec:7.3} and appendix~\ref{sec:C}). Thus, at the point $c_{s}
=c_{s,1} =1-\textrm{const}\;\phi _{0} $, the topological transition takes place at which the
heterophase correlated domains  transform into homophase ones. It is important
that this transition does not presuppose the Fischer cluster equilibration on
scales $\xi _\textrm{FC} >\xi _{fl} $. This result denotes a change of the structural relaxation
mode at glass transition  considered in \cite{76}.

 The mesoscopic theory of thermodynamics and dynamics of the glass-forming liquids
and glasses is connected with the microscopic approach based on the consideration of the
potential energy landscape (see \cite{79} and references cited) by the landscape coarsening
procedure used while deducing the efficient Hamiltonian (appendix~\ref{sec:A}). The coefficients
of the fluctuon interaction save the memory on the microscopic potential energy
landscape.

\appendix

\section{The bounded phase space and efficient mesoscopic Hamiltonian}
\label{sec:A}

Below $T_{m} $, a crystalline state is the most probable one. It occupies a phase
space region $\Omega _\textrm{cr} $ of the total phase space $\Omega $.  $\Omega $ can be
presented as the sum of the regions belonging to crystalline and non-crystalline states,
\begin{equation}
\Omega
=\Omega _\textrm{cr} +\Omega _\textrm{ncr}\,.
\label{A.1}
\end{equation}
 Excluding $\Omega _\textrm{cr} $, we obtain the bounded phase space belonging to non-crystalline
states. The bounded partition function
\begin{equation}
\stackrel{\frown}{Z}\left(P,T\right)=\int _{\Omega _\textrm{ncr} }\exp \left[-E_{N}
\left(x,p\right)\beta \right]\rd^{3N}x \;\rd^{3N} p
\label{A.2}
\end{equation}
determines the free energy of the non-crystalline state
\begin{equation}
G\left(P,T\right)=-T\ln \stackrel{\frown}{Z}\left(P,T\right).
\label{A.3}
\end{equation}
In \cite{36}, the procedure of derivation of the equation of the free energy  in terms
of the order parameter \eqref{GrindEQ__3_}, $G\left(\left\{c(x)\right\}\right)$,
(it is called the efficient  Hamiltonian in \cite{36}) is expounded. It is based on the map
of the phase space on the functional space of the order parameter $\left\{c\left(x
\right)\right\}$,
\begin{eqnarray}
G\left(\left\{c(x)\right\}\right)=-T\ln \int _{\Omega _\textrm{ncr} }\exp \left[-E_{N}
\left(x,p\right)\beta \right]\prod _{\vec{q},i}\delta \big(c_{i} (\vec{q})-c_{i}
(\vec{q};x,p)\big) \rd^{3N}x \;\rd^{3N} p.
\label{A.4}
\end{eqnarray}
Here, $c_{i} \left(\vec{q};x,p\right)$ are Fourier transforms of the components of
the order parameter at a fixed coordinate $(x,p)$ of  the bounded 6\textit{N}-dimensional
phase space $\Omega _\textrm{ncr} $.

Performing the integration in the functional space, we have
\begin{equation}
G\left(P,T\right)=-T\ln \int \exp \left[-G\left(\left\{c(\vec{q})\right\}\right)
\beta \right] \prod _{\vec{q},i}\rd c_{i} \left(\vec{q}\right).
\label{A.5}
\end{equation}
Selecting in this expression the contribution of long-wavelength components, $q\to
0$, and performing a  polynomial expansion in powers of  $c_{i} \left(x\right)$ \eqref{GrindEQ__7_}--\eqref{GrindEQ__12_}, we obtain
\begin{eqnarray}
G\left(P,T\right)&=&-T\ln \int \exp \left[-H_\textrm{eff} \left(\left\{c
\right\}\right)\beta \right] \prod _{qr_{0} \ll 1,i}\rd c_{i} \left(\vec{q}\right) +\tilde{G}(P,T),
\nonumber
\\
H_\textrm{eff} \left(\left\{c\right\}\right)&=&G\left(P,T;\left\{c(x)\right\}\right).
\label{A.6}
\end{eqnarray}
 The term $\tilde{G}(P,T)$ takes into account the spatial fluctuations of the order
parameter with $qr_{0} \sim 1$. To include this summand into consideration is important
in the vicinity of critical points. It generates random fields and has an impact on
criticality. It is shown in \cite{10} that at the end point on the phase coexistence curve,
the first order phase transition can take place due to the impact of the random field.

Within
the framework of the method of cooperative variables, used while considering
the gas-liquid critical point \cite{80} and systems with the Ising-type Hamiltonian
\cite{81}, a  procedure  of accounting of $\tilde{G}(P,T)$ in the vicinity of the critical
point is expounded.

\section{Solutions of the equations of state }
\label{sec:B}

To get solutions of the equations \eqref{GrindEQ__19_}--\eqref{GrindEQ__22_},
\eqref{GrindEQ__32_}, \eqref{GrindEQ__34_},  let us consider the solutions of  equations~\eqref{GrindEQ__32_}, \eqref{GrindEQ__34_} taking  $c_{s} \left(T\right)$ as an unknown
function which should be determined later on using equations \eqref{GrindEQ__20_}--\eqref{GrindEQ__22_}.

If $s_{1}
>s_{2} $, the ``external field'' $h_{12} $ is positive below $T_{12} $ and negative
above $T_{12} $.  The  first order phase transition takes place and  discontinuous
transformation of the phase 1 into phase 2 takes place at $T_{12} $ if
\begin{equation}
c_{s} \left(T_{12} \right)\tilde{g}_{12} >2T_{12}\,.
\label{B.1}
\end{equation}
Under this condition, the solution similar to the solution \eqref{GrindEQ__29_} of
equation~\eqref{GrindEQ__20_} exists but it is unstable.

To get the other two solutions near $T_{12} $, equation~\eqref{GrindEQ__32_}  can be rewritten
taking into account the relation \eqref{GrindEQ__34_} as follows:
\begin{eqnarray}
-2\alpha ^{*} c_{s} \tilde{g}_{12} +T\ln \frac{1+2\alpha ^{*} }{1-2\alpha ^{*} }
\approx  2\left(-c_{s} \tilde{g}_{12} +2T\right)\alpha ^{*} +\frac{16}{3} T\left(
\alpha ^{*} \right)^{3}
=\Delta s_{12} \left(T-T_{12} \right).
\label{B.2}
\end{eqnarray}
Here, $\alpha ^{*} =c\frac{1}{2}$ $\alpha ^{*} =\frac{1}{2} -c_{1}^{*} =c_{2}^{*}
-\frac{1}{2}$.

Near $T_{12} $, as it follows from (\ref{B.2}), there exist two other stable solutions,
\begin{equation}
\alpha _{1,2}^{*} =\pm \frac{\sqrt{3} }{2} \left(\frac{c_{s} \tilde{g}_{12} -2T}{2T}
\right)^{1/2} \left(1-\frac{h_{12} }{c_{s} \tilde{g}_{12} -2T} \right)
\label{B.3}
\end{equation}
at $\left(\alpha ^{*} \right)^{2} \ll \left(c_{s} \tilde{g}_{12} -2T\right)T^{-1} $.

Equation
\begin{equation}
c_{s} \left(T\right)\tilde{g}_{12} =2T
\label{B.4}
\end{equation}
determines  critical  temperatures of the system. Near a critical point, where
$\left(\alpha ^{*} \right)^{2} \geqslant \left(c_{s} \tilde{g}_{12} -2T\right)T^{-1} $,
as it follows from (\ref{B.2})
\begin{equation}
\alpha ^{*} \approx \frac{1}{2} \left[\frac{3\Delta s_{12} \left(T-T_{12} \right)}{2T}
\right]^{1/3}.
\label{B.5}
\end{equation}
Turning to the search of self-consistent solutions $c_{s} (T)$, one can use the expressions
found in subsection~\ref{sec:8.1} and  (\ref{B.3}), (\ref{B.5}).  At  $T>T_{e} $, equation~\eqref{GrindEQ__23_} describes
the required solutions $c_{s,1} (T)$ and $c_{s,2} (T)$ if we put correspondingly
\begin{equation}
c_{1}^{*}
=c_{1,1}^{*} \left(T\right)=\frac{1}{2} +\alpha _{1}^{*} \left(T\right) \qquad \text{or}
\qquad
c_{1}^{*}
=c_{1,2}^{*} \left(T\right)=\frac{1}{2} +\alpha _{2}^{*} \left(T\right).
\label{B.6}
\end{equation}
Graphic
representation of the solutions $c_{1,1}^{*} \left(T\right)$, $c_{1,1}^{*} \left(T
\right)$ and  $c_{s,1} (T)$, $c_{s,2} (T)$ is shown in figure~\ref{fig3}.

The jump of $c_{1}^{*} $ at $T=T_{12} $ is $\Delta c_{1}^{*} \left(T_{12} \right)
\sim 1$. Hence, in accordance with \eqref{GrindEQ__23_}
\begin{eqnarray}
\Delta c_{s} \left(T_{12} \right)=c_{s,2} \left(T_{12} \right)-c_{s,1} \left(T_{12}
\right)\approx \left(s_{1} -s_{2} \right)k_{0}^{-1}c_{s} \left(T_{12} \right).
\label{B.7}
\end{eqnarray}
Here, $s_{1} ,s_{2} $ are entropies of  $s$-fluctuons.

The heat of the phase transition is equal to
\begin{eqnarray}
\Delta H&=&\Delta c_{s} \left(T_{12} \right)H_{fs}  +c_{s}\left(T_{12} \right)H_{12} \simeq c_{s} \left(T_{12} \right)\left(\frac{s_{1}-s_{2}}{k_{0}}H_{fs} +H_{12}\right)
\nonumber \\
&= &c_{s} \left(T_{12}\right)\frac{s_{1}-s_{2}}{k_{0}}
\left(H_{fs} +H_{12} \right),
\label{B.8}
\end{eqnarray}
$H_{sf} $ is the heat of the fluid-solid phase transition and $H_{12}= \left(s_{1} -s_{2} \right)k_{0}^{-1}T_{12}$ is
the heat of  $1\leftrightarrow 2$ solid-solid phase transition.

\section{Thermodynamics and structure of the Fischer cluster}
\label{sec:C}
The contribution of volumetric interactions into the free energy density, as seen
from  \eqref{GrindEQ__10_}, is as follows:
\begin{equation}
g_{V} (P,T)=2\pi \int \Phi (r)w_{ss} (r)r^{2} \rd r', \qquad
r=\left|x-x' \right|,
\label{C.1}
\end{equation}
\begin{equation}
w_{ss}(r)= \langle c_{s} (x)c_{s} (x')\rangle  = c_{_{S} }\omega _{ss} (r)+c_{s}^{2}\,,
\label{C.2}
\end{equation}
where $c_{s} $ is the mean value of $c_{s} (x)$.

Assuming that the $s$-fluctuons form fractal aggregates of dimension $D_{f} $ with
correlation length $\xi _\textrm{FC} $, we look for the correlator $w_{ss} (r)$of the
following form
\begin{equation}
w_{ss} (r)=c_{s} \left(\frac{r_{0} }{r} \right)^{3-D_{f} } \exp (-r/\xi _\textrm{FC} )+c_{s}
^{2} \left[1-\exp (-r/\xi _\textrm{FC} )\right], \qquad r\geqslant
r_{0}, \qquad  1\leqslant D_{f} \leqslant 3.
\label{C.3}
\end{equation}
The
condition $1\leqslant D_{f} \leqslant 3$ provides the topological connectivity of the Fischer
cluster. It follows from (\ref{C.2}), (\ref{C.3}) that
\begin{equation}
\omega _{ss} =\left[\left(\frac{r_{0} }{r} \right)^{3-D_{f} } -c_{s} \right]\exp
(-r/\xi _\textrm{FC} ), \qquad
r\geqslant r_{0}\,.
\label{C.4}
\end{equation}
The parameters $D_{f} $ and $\xi _\textrm{FC} $ should be found minimizing the free energy
\eqref{GrindEQ__7_}.

As the first step, we  find the chemical potential $\overline{\mu }_{s} \left(P,T
\right)$ of the reference system accounting for the volumetric interactions of non-correlated
fluctuons, at  $\xi _\textrm{FC} \to 0$
\begin{eqnarray}
\overline{\mu }_{s} \left(P,T\right)&=&g_{s}^{0} \left(P,T\right)+g_{ss} \left(P,T
\right)c_{s} +T\ln c_{s} +\varphi _{0} c_{s}\,, \label{C.5}\\
\varphi _{0}& =&4\pi \int \Phi (r)r^{2} \rd r' =-4\pi R_{0} ^{2} \varphi.
\label{C.6}
\end{eqnarray}
Hence, the equilibrium equation reads
\begin{equation}
(1-2c_{s} )\tilde{g}_{sf,v} +T\ln \frac{c_{s} }{1-c_{s} } =h_{sf,v}\,,
\label{C.7}
 \end{equation}
\begin{eqnarray}
 \tilde{g}_{sf,v} &=& g_{sf}(P,T)-g_{ss} (P,T)/2-\varphi _{0} /2, \nonumber \\
h_{sf,v} &=& g_{f}^{0}
(P,T)-g_{s}^{0} (P,T)-g_{ss} (P,T)/2-\varphi _{0} /2.
\label{C.8}
\end{eqnarray}
Thus, with $\omega _{ss} (r)=0$, the role of volumetric interaction is reduced to a
renormalization of coefficients of the equilibrium equation. If $\tilde{g}_{sf,v}
<2T_{e} $,  $c_{s} $ increases continuously with the temperature decrease (see
subsection~\ref{sec:7.1}). Just this case is considered hereinafter.

Let us denote by $g^{0} (c_{s} )$ the free energy of $s$-fluctuon in equilibrium liquid
at $\omega _{ss} (r)=0$,  and by $g_\textrm{FC} \left(c_{s} ;D_{f} ,\xi _\textrm{FC} \right)$ its
value at $\omega _{ss} (r)\ne 0$. The difference of these quantities determines the
correlation free energy, $g_\textrm{corr} \left(c_{s} ;D_{f} ,\xi _\textrm{FC} \right)$, as function
of $c_{s} $,
\begin{equation}
g_\textrm{corr} \left(c_{s} ;D_{f} ,\xi _\textrm{FC} \right)=g_\textrm{FC}
\left(c_{s} ;D_{f} ,\xi _\textrm{FC} \right)-g^{0} (c_{s} ).
\label{C.9}
\end{equation}
It follows from (\ref{C.1}) that
\begin{eqnarray}
&&\hspace{-8mm}g_\textrm{corr} (c_{s} ;D_{f} ,\xi _\textrm{FC} )=2\pi \int _{r_{0} }^{\infty }\Phi (r)\omega _{ss}
(r)r^{2} \rd r' -c_{s} \Delta s_\textrm{corr} T \nonumber \\
&&=-2\pi R_{0} ^{2} c_{s}^{2} \left(\frac{\xi _\textrm{FC} }{R_{0} +\xi _\textrm{FC} }
\right)^{\alpha } \left[\Gamma \left(\alpha ,y\right)\frac{c_{s,\textrm{CD}} }{c_{s} } -\Gamma
\left(2,y\right)\left(\frac{\xi _\textrm{FC} }{R_{0} +\xi _\textrm{FC} } \right)^{2-
\alpha } \right]\varphi -c_{s} \Delta s_\textrm{corr} T \nonumber \\
&&=
\frac{1}{2} c_{s}^{2} \left(\frac{\xi _\textrm{FC} }{R_{0} +\xi _\textrm{FC} } {\kern 1pt} \right)^{
\alpha } \left[\Gamma \left(\alpha ,y\right)\frac{c_{s,\textrm{CD}} }{c_{s} } -\Gamma \left(2,y
\right)\left(\frac{\xi _\textrm{FC} }{R_{0} +\xi _\textrm{FC} } {\kern 1pt} \right)^{2-\alpha }
\right]\varphi _{0} -c_{s} \Delta s_\textrm{corr} T,
\label{C.10}
\end{eqnarray}
$\Delta s_\textrm{corr} $ is the entropy difference per $s$-fluctuon due to the correlation,
\begin{eqnarray}
\Delta s_\textrm{corr}& =&-4\pi N_{\xi }^{-1} \int _{r_{0} }^{\xi }\Big(\big\{c_{s} (r)
\ln c_{s} (r)+\left[1-c_{s} (r)\right]\ln \left[1-c_{s} (r)\right]\big\} \nonumber \\
&&
-\left[c_{s}
\ln c_{s} +\left(1-c_{s} \right)\ln \left(1-c_{s} \right)\right]\Big) r^{2} \rd r, \nonumber \\
c_{s} (r)&=&r^{D_{f} -3}, \qquad
N_{\xi } =4\pi c_{s}\xi _\textrm{FC}^{3}/3r_{0}^{-3}   =\frac{4\pi }{3} \left(\xi _\textrm{FC} /r_{0} \right)^{D_{f}},
\label{C.11}
\end{eqnarray}
$N_{
\xi } $ is the number of $s$-fluctuons in the correlated part of the fractal; $\Gamma
\left(\alpha ,y\right)$ is an incomplete gamma-function, $\alpha =D_{f} -1>0$, $y=r_{0}
/\xi _{fl} $,
\begin{equation}
c_{s,\textrm{CD}} =\left(\frac{r_{0} }{\xi _{fl} } \right)^{3-D_{f} }
\label{C.12}
\end{equation}
is the mean fraction of $s$-fluctuons within the correlated domain.

Integration in (\ref{C.11}) at $\overline{c}_{s} \ll 1$ gives
\begin{equation}
\Delta s_\textrm{corr} =-\frac{3-D_{f} }{D_{f} } -\ln \frac{3}{D_{f} }\,.
\label{C.13}
\end{equation}
The correlation length $\xi _\textrm{FC} $ can be estimated as follows. Noting, that in
a fractal of dimension $D_{f}$ and radius $\xi _\textrm{FC} $
\begin{equation}
c_{s} (\xi )\approx \left(\frac{\xi }{r_{0} } \right)^{D_{f} -3},
\label{C.14}
\end{equation}
we have
\begin{equation}
\xi _\textrm{FC} \left(c_{s} ,D_{f} \right)\approx r_{0} \left(c_{s} \right)^{\frac{1}{D_{f}
-3} }.
\label{C.15}
\end{equation}
The fractal dimension $D_{f} $ is determined by relation (\ref{C.12}) if equilibrium
value of $c_{s,\textrm{CD}} $ is known. To find it, let us consider the free energy of $s$-fluctuon
within CD as a function of $c_{s,\textrm{CD}} $ and minimize it.  Denoting it by $g_\textrm{CD} \left(c_{s}
,_\textrm{CD} \right)\equiv g_\textrm{CD} \left(c_{s} ;D_{f} ,\xi _\textrm{FC} \right)$ , we have from
(\ref{C.9})
\begin{equation}
g_\textrm{CD} (c_{s,\textrm{CD}} )=g^{0} (c_{s,\textrm{CD}} )+g_\textrm{corr} (c_{s,\textrm{CD}} ;D_{f} ,\xi _\textrm{FC} ).
\label{C.16}
\end{equation}
Minimum
of $g_\textrm{CD} (c_{s,\textrm{CD}} )$ is attained at the value $c_{s,\textrm{CD}} $ being the solution of
the equation
\begin{equation}
\partial g_\textrm{CD} (c_{s,\textrm{CD}} )/\partial c_{s,\textrm{CD}} =0
\label{C.17}
\end{equation}
under the condition
\begin{equation}
\partial ^{2} g_{s,\textrm{CD}} (c_{s,\textrm{CD}} )/\partial c_{s,\textrm{CD}}^{2} >0.
\label{C.18}
\end{equation}
Noting that $g^{0} (c_{s,\textrm{CD}} )$ has a minimum at $c_{s,\textrm{CD}} =c_{s} $ and expanding
(\ref{C.16}) in series on degrees of $\delta c_{s} =c_{s,\textrm{CD}} -c_{s} $, we have
\begin{eqnarray}
g_\textrm{CD} (c_{s,\textrm{CD}} )&=&\frac{1}{2} \frac{\partial ^{2} g^{0} (c_{s}
)}{\partial c_{s}^{2} } \delta c_{s} ^{2}
- \frac{1}{2} \Gamma \left(\alpha ,y
\right)\left(\frac{\xi _\textrm{FC} }{R_{0} +\xi _\textrm{FC} }  \right)^{\alpha } \left(c_{s}^{2} +2c_{s} \delta c_{s} +\delta c_{s} ^{2}\right)
\varphi _{0}\nonumber \\
&&
 -\;\Gamma \left(2,y\right)\left(\frac{\xi _\textrm{FC} }{R_{0} +\xi _\textrm{FC} } \right)^{2} \varphi _{0} c_{s}^{2}\,.
\label{C.19}
\end{eqnarray}
It follows from equation (\ref{C.17})--(\ref{C.19}) that
\begin{equation}
\delta c_{s} =\frac{I(D_{f} ,\xi )c_{s} }{\partial ^{2} g^{0} \left(c_{s} \right)/
\partial c_{s}^{2} -I(D_{f} ,\xi )}\,,
\label{C.20}
\end{equation}
where
\begin{equation}
I(D,\xi )={\kern 1pt} {\kern 1pt} \frac{1}{2} \Gamma \left(\alpha ,y\right)\left(
\frac{\xi _\textrm{FC} }{R_{0} +\xi _\textrm{FC} } {\kern 1pt} \right)^{\alpha } \varphi _{0}\,.
\label{C.21}
\end{equation}
Thus,
\begin{equation}
c_{s,\textrm{CD}}
=\left[1+\frac{I(D_{f} ,\xi _\textrm{FC} )}{\partial ^{2} g^{0} \left(c_{s} \right)/\partial
c_{s}^{2} -I(D_{f} ,\xi _\textrm{FC} )} \right]c_{s}\,.
\label{C.22}
\end{equation}
It is seen that $\delta c_{s} \sim \phi _{0} $ and $c_{s,\textrm{CD}} \geqslant \overline{c}_{s} $ due
to the condition (\ref{C.18}).

Equations (\ref{C.12}) and (\ref{C.22}) determine the fractal dimension. Equation (\ref{C.12}) gives
\begin{equation}
D_{f} =3-\left(\ln y\right)^{-1} \ln c_{s,\textrm{CD}}\,,
\qquad
 y=r_{0} /\xi _{fl}\,.
\label{C.23}
\end{equation}
Since $1\leqslant D_{f} \leqslant 3$,
\begin{equation}
2\geqslant \left(\ln y\right)^{-1} \ln c_{s,\textrm{CD}} \geqslant 0.
\label{C.24}
\end{equation}
As it follows from (\ref{C.22}) and (\ref{C.24}), the fractal dimension changes within the
range $1\leqslant D_{f} \leqslant 3$ when $c_{s} $ changes  within the range
\begin{equation}
c_{s,0} \leqslant c_{s} \leqslant c_{s,1}\,,
\label{C.25}
\end{equation}
where
\begin{eqnarray}
c_{s,0}&=&y^{2} \left[1+\frac{2I(D_{f} ,\xi )}{\partial ^{2} g^{0} \left(c_{s} \right)/\partial
c_{s}^{2} -2I(D_{f} ,\xi )} \right]^{-1} >0, \nonumber \\
c_{s,1}& =&\left[1+\frac{2I(D_{f} ,\xi )}{\partial ^{2} g^{0} \left(c_{s} \right)/
\partial c_{s}^{2} -2I(D_{f} ,\xi )} \right]^{-1} \approx 4c_{s,0} <1.
\label{C.26}
\end{eqnarray}
 As it follows from (\ref{C.26}), $c_{s,0} <0.16$ with $y=r_{0} /\xi _{fl} \approx 0.4$,
i.e.,   $c_{s,0} $ is nearly equal to the percolation threshold of the solid-like
fraction. The upper bound of the $c_{s} $-range in which the Fischer cluster exists,  $c_{s,1} $,
decreases $\sim \phi _{0} $ with an increase of the $s$-fluctuons gravitation strength $\phi
_{0} $.  It is worth to note that $D_{f} \to 3$ and $\xi _\textrm{FC} \to \infty $ when $c_{s} $ approaches $c_{s,1} $ from below. It
means that with $c_{s} \geqslant c_{s,1} <1$, the solid-like fraction consists of the connected
3-dimensional solid-like clusters of size $r>\xi _{fl} $.

\clearpage

\ukrainianpart

\title{Гетерофазні стани рідини: термодинаміка, структура, динаміка
}
\author{О.С. Бакай}
\address{ННЦ Харківський фізико-технічний інститут, 61108 Харків, Україна}

\makeukrtitle

\begin{abstract}
\tolerance=3000%
Представлено огляд теоретичних результатів та експериментальних даних щодо термодинаміки, структури і динаміки
гетерофазних склоутворювальних рідин.  Теоретичний підхід базується на моделі мезоскопічних гетерофазних флуктуацій, яка була розвинута
в рамках підходу обмеженої статистичної функції.
Розглянуто явище кластер Фішера, перехід у фазу скла,  перетворення рідина-рідина, параметричну фазову діаграму, колективну динаміку і
фрагільність склоутворювальних рідин.
\keywords склоутворювальні рідини, перехід у фазу скла, кластер Фішера, поліморфізм, параметрична фазова діаграма
\end{abstract}

\end{document}